\documentclass[aps, prl, twocolumn]{revtex4-1}
\usepackage[british]{babel}
\usepackage{amsmath,amssymb,graphicx,xcolor,hyperref}

\newcommand{\red}[1]{#1}
\newcommand{\changes}[1]{#1}

\begin{document}

\title{Universality of entanglement transitions from stroboscopic to continuous measurements}

\author{M. Szyniszewski}
\email{mszynisz@gmail.com}
\affiliation{Department of Physics, Lancaster University, Lancaster LA1 4YB,
  United Kingdom}
\affiliation{Department of Physics
and Astronomy, University College London, London WC1E 6BT, United Kingdom}

\author{A. Romito}
\affiliation{Department of Physics, Lancaster University, Lancaster LA1 4YB,
  United Kingdom}

\author{H. Schomerus}
\affiliation{Department of Physics, Lancaster University, Lancaster LA1 4YB,
  United Kingdom}

\begin{abstract}
  Measurement-driven transitions between extensive and sub-extensive scaling of
  the entanglement entropy receive interest as they illuminate the intricate
  physics of thermalization and control in open interacting quantum systems.
  Whilst this transition is well established for stroboscopic measurements in
  random quantum circuits, a crucial link to physical settings is its extension
  to continuous observations, where for an integrable model it has been shown
  that the transition changes its nature and becomes immediate. Here, we
  demonstrate that the entanglement transition at finite coupling persists if
  the continuously measured system is randomly nonintegrable, and show
  that it is smoothly connected to the transition in the stroboscopic models.
  This provides a bridge between a wide range of experimental settings and the wealth of
  knowledge accumulated for the latter systems.
\end{abstract}

\maketitle

Subjecting a complex quantum system to observations can have drastic effects on
its time evolution. The most celebrated example is the quantum Zeno
effect~\cite{Degasperis1974, Misra1977, Peres1980}, according to which
\emph{continuous} projective measurements can freeze the dynamics of a quantum
system completely. Recent work has established~\cite{Li2018, Chan2018,
Skinner2018, Li2019, Szyniszewski2019} and developed~\cite{Jian2020, Li2020,
Chen2020, Choi2019, Gullans2019purification, Kuo2019, Nahum2019, Roy2019,
Zabalo2020, Zhang2020, Fan2020, Bao2020, Bera2020, Lavasani2020, Sang2020,
Ippoliti2020, Tang2020, Rossini2020, Goto2020, Fuji2020, Snizhko2020,
Gebhart2020, Gullans2019, LopezPiqueres2020, Shtanko2020, Noh2020, Fang2020,
Alberton2020, Lunt2020, Turkeshi2020, Li2020correcting} an
illuminating extension of this effect, where the quantum dynamics change in a
phase transition when \emph{stroboscopic} measurements occur with sufficient
strength or frequency. This transition is manifested in the entanglement
characteristics of the system, as captured by the entanglement entropy
\begin{equation}
  S=\mathrm{tr}\,(\rho_A\ln\rho_A)
\end{equation}
with the reduced density matrix of a subsystem $A$. In the transition, the
entropy changes its scaling with the system size~\cite{Znidaric2008, Eisert2010,
Bardarson2012, Bauer2013, Kjall2014, Luitz2015, Chan2018} from extensive,
indicating ergodic many-body dynamics, to sub-extensive, signaling localization
of the underlying quantum-coherent correlations.

A key question to make this rapidly growing body of knowledge on stroboscopic
systems  applicable to physical settings is the fate of the entanglement
transition for continuous variable-strength observations. These not only more
accurately reflect the reality of many experimental
architectures~\cite{petta2005coherent, kim2014quantum, west2019gate,
Murch2013observing, neumann2010single, Aasen2016milestones, manousakis2020weak,
steiner2020readout, munk2020parity, maioli2005nondestructive}, but also enable to apply this knowledge to the
generic effects of coupling to an environment that may not per se have been
designed to carry out a measurement. For an integrable system, it has been shown
that the transition can indeed completely change its nature when observations
become continuous, in that it then can occur at infinitesimal small measurement
strengths~\cite{Cao2019}.

Here, we show that the transition from the stroboscopic models is reinstated for
continuous observations of a randomly evolving, nonintegrable, system. We
achieve this by formulating a model that allows us to interpolate between a
stroboscopic random circuit and a continuously evolving one, and trace the
entanglement characteristics numerically in terms of the entanglement entropy
and mutual information. The established link between these limits lends further
relevance to  deep results arrived for the stroboscopic circuits---such as
emerging conformal symmetry~\cite{Skinner2018, Li2019, Jian2020, Li2020,
Chen2020} as well as approximations that permit to reach very large system
sizes~\cite{Chan2018, Li2018, Li2019, Gullans2019, Napp2019, Zhou2019,
Shtanko2020, Noh2020, LopezPiqueres2020}---giving them direct bearing on a much
wider range of experimental settings.

\begin{figure}[b]
  \centering
  \includegraphics[width=\linewidth]{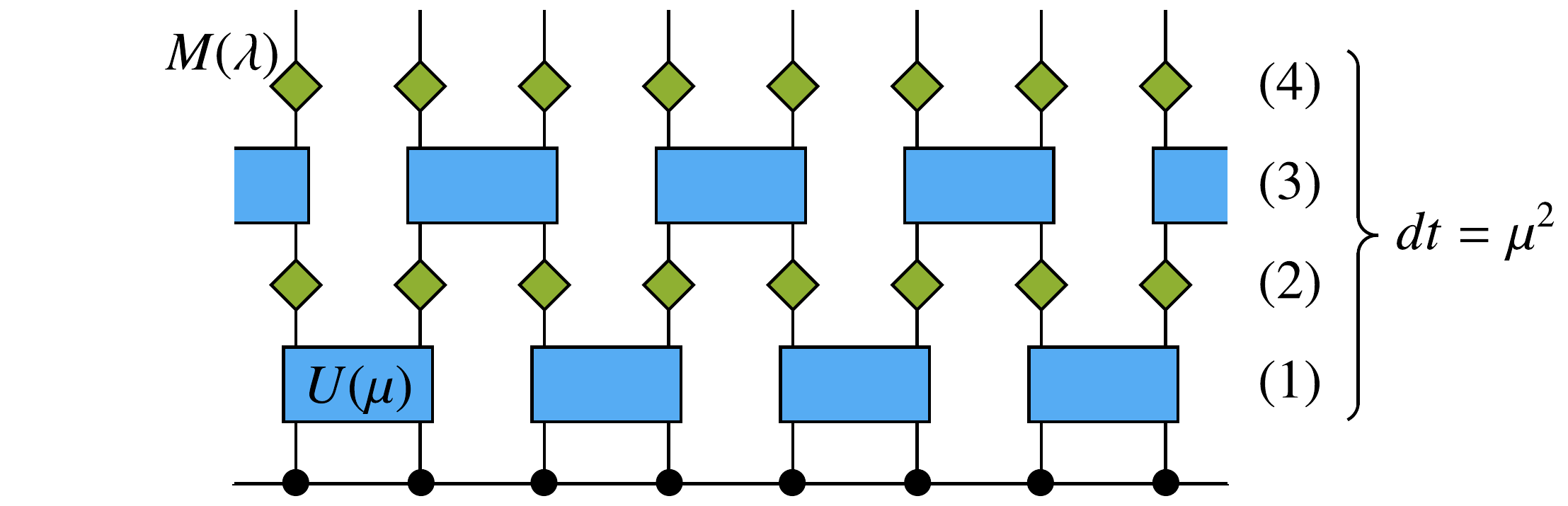}

  \caption{We study the entanglement dynamics in a random circuit model, combining unitary evolutions
   $U$ and measurements $M$ such that one can interpolate between the continuum limit ($U$ near the identity matrix and measurements weak) and widely studied fully random, stroboscopic models. This is achieved by equipping the unitary matrices with a parameter $\mu$ that determines the physical time scale of the dynamics according to $dt\sim \mu^2$, and the measurements with a parameter $\lambda$ so that the effective measurement strength is given by $\lambda_0=\lambda/\mu$.
   \label{fig:circuit}}
\end{figure}

\emph{Model}.---We base our modeling on the universal quantum-circuit
architecture~\cite{Nahum2017, Keyserlingk2018, Nahum2018, Chan2018} shown in
Fig.~\ref{fig:circuit}, which describes the dynamics of $L$ spins (dots)
evolving under the action of unitary gates $U$ (rectangles) and nonunitary
measurement operations $M$ (diamonds) on individual spins. Two layers of gates
and measurements make up one time step, $dt$, and iteration over $n$ steps
induces a discrete time evolution of the quantum state $|\psi_n\rangle$. In the
original design~\cite{Li2018, Chan2018, Skinner2018}, the gates are completely
random, according to unitary matrices $U$ drawn from a circular ensemble with
probability distribution given by the corresponding Haar measure, whilst the
measurements are projective, so that the time step $dt=O(1)$ in terms of
physical times scales governing the dynamics. This design can be easily adapted
to other situations, including systems with deterministic
dynamics~\cite{Cao2019, Tang2020, Rossini2020, Fuji2020, Goto2020,
Ippoliti2020, Alberton2020} or other types of
measurements~\cite{Szyniszewski2019, Sang2020, Lavasani2020, Chen2020,
Fuji2020, Alberton2020}.

Here, we carry out two such modifications designed to change the dynamics and observation strength over the time scale $dt$, thereby allowing us to take the continuum limit in which $dt\to 0$.

(A) The unitary matrices $U$ are generated from an ensemble parameterized by
$0\leq \mu\leq 1$, which interpolates between matrices close to the identity
matrix ($\mu\ll 1 $) and the exact Haar measure ($\mu=1$). This is realized
using the Poisson kernel~{\cite{Hua1963}},
\begin{equation}
  U = \red{(V+ \sqrt{1-\mu^2}\openone)(\openone+\sqrt{1-\mu^2}V)^{-1}},
\end{equation}
where $V$ is a random unitary matrix distributed according to the Haar
measure. The
latter is recovered for $\mu=1$, where $U=V$. For $\mu\ll 1$, the matrices
localize close to the identity matrix,
\begin{equation}
  U \approx \openone -i\,dt\,H_\mathrm{eff},\quad dt\equiv \mu^2,\quad
  H_\mathrm{eff}=\frac{i}{2}\red{(V-\openone)(V+\openone)^{-1}},
\end{equation}
which identifies the Cayley transform of $V$ as
the effective Hamiltonian $H_\mathrm{eff}$, and sets the physical time scale
according to the resulting stable Cauchy process \footnote{The Cauchy process
arises from the  Lorentzian distribution of $H_\mathrm{eff}$; see, e.g.,
Ref.~\cite{Brouwer1995}. An analogous entanglement dynamics can be obtained
from a Wiener process where $H_\mathrm{eff}$ is taken from the Gaussian unitary
ensemble with matrix elements $H_{\mathrm{eff},{lm}}=O(1/\mu)$; \red{for more details see \cite{Note2}.}}.  However, this
Wiener process does not permit the exact extrapolation to the stroboscopic
case.  Between these
limits, the matrices
preserve unitarity and maintain  randomness according to a probabilistic
maximal-entropy principle \cite{Mello1985, Brouwer1995}.

(B) The projective measurements are replaced by weak measurements\red{,
implemented by coupling the $z$\nobreakdash-component of a given spin to an external
pointer with a continuous readout $x$, prepared initially in a Gaussian state.
Measurements are} of strength
$\lambda$, ranging from the case of no measurements $(\lambda = 0)$ to the
standard case of projective measurements $(\lambda \rightarrow \infty)$. These
take the form of positive-operator-value measurements \cite{Jacobs2014quantum,
Wiseman2009quantum} with Kraus operators
\begin{equation}
  M(x)=G (x - \lambda) \Pi^+_i + G (x + \lambda) \Pi^-_i, \label{eq:kraus}
\end{equation}
where $G (x) = \exp (- x^2 / 2) / \pi^{1 / 4}$ is a Gaussian of unit width
centered around zero, and $\Pi^{\pm}_i = (1 \pm \sigma^z)_i/2$ are projection
operators onto spin-up or spin-down on site $i$. For a given readout $x$, the
system state is updated via
\begin{equation}
  \vert \psi \rangle \to \frac{1}{\sqrt{P(x)}} M(x) \vert \psi \rangle,
  \label{eq:update}
\end{equation}
where $P(x)=\langle \psi |M(x)^\dagger M(x) |  \psi \rangle$ is the probability
distribution of the measurement output. For small $\lambda$, the measurement model reduces
to a generic Wiener process
\begin{equation}
  \vert \psi \rangle\to\mathcal{N} \left[ 1  - \sum_i \left( \lambda^2 \langle
  \sigma^z_i \rangle + \delta W_{i} \right) \sigma^z_i \right] \vert \psi
  \rangle, \label{eq:stochastic}
\end{equation}
where the random variables  $W_{i}$ are independently Gaussian-distributed with
zero mean and variance $\lambda^2$, and $\mathcal{N}$ is a normalization constant.

\begin{figure}[t]
  \centering
  \includegraphics[width=0.95\linewidth]{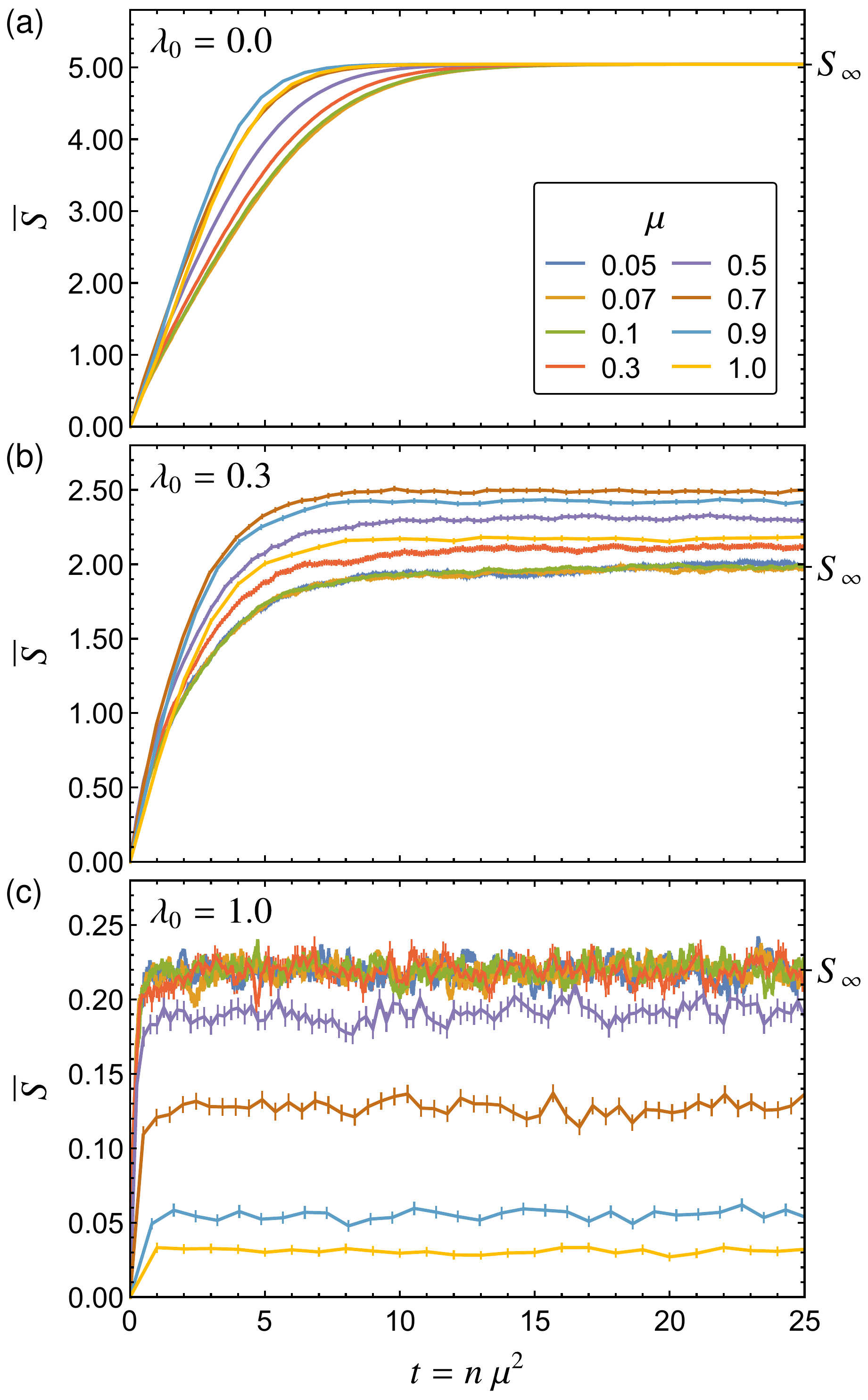}

  \caption{Entanglement dynamics in the random-circuit model of
  Fig.~\ref{fig:circuit}, as captured by the time evolution of the averaged
  bipartite entanglement entropy $\overline S$. Time is measured in units
  $t=n\mu^2$, the system size is $L = 16$, and results are averaged over 1000
  realizations. The different panels fix the effective measurement strength to
  (a) $\lambda_0 = 0$, (b) $\lambda_0 = 0.3$, and (c) $\lambda_0 = 1.0$, with
  the different curves corresponding to different choices of $\mu$. Throughout the whole dynamics, the curves
  collapse for $\mu\lesssim 0.1$,
  which indicates entering the continuum regime. Increasing the measurement strength suppresses
  the quasistationary value $S_\infty$, which raises the question of an
  entanglement transition addressed in the subsequent figures. \label{fig:SvsTime}}
\end{figure}

Writing the intrinsic scale of this process as $\lambda^2 = \lambda_0^2\,dt =
\lambda_0^2\mu^2$, the effective strength of the measurement in our model is
therefore given by
\begin{equation}
  \lambda_0=\lambda/\mu,
\end{equation}
which has to be kept fixed as we send $dt=\mu^2\to 0$. The physical time scale
for the dynamics is then given by $t=n\,dt=n\,\mu^2$, where $n$ is the number of
steps through the circuit depicted in Fig. ~\ref{fig:circuit}. Our main result
will be to establish that an entanglement transition occurs at a finite value of
$\lambda_0$, for all scenarios from the continuum limit to the fully random
stroboscopic case.

\begin{figure}[t]
  \centering
  \includegraphics[width=\linewidth]{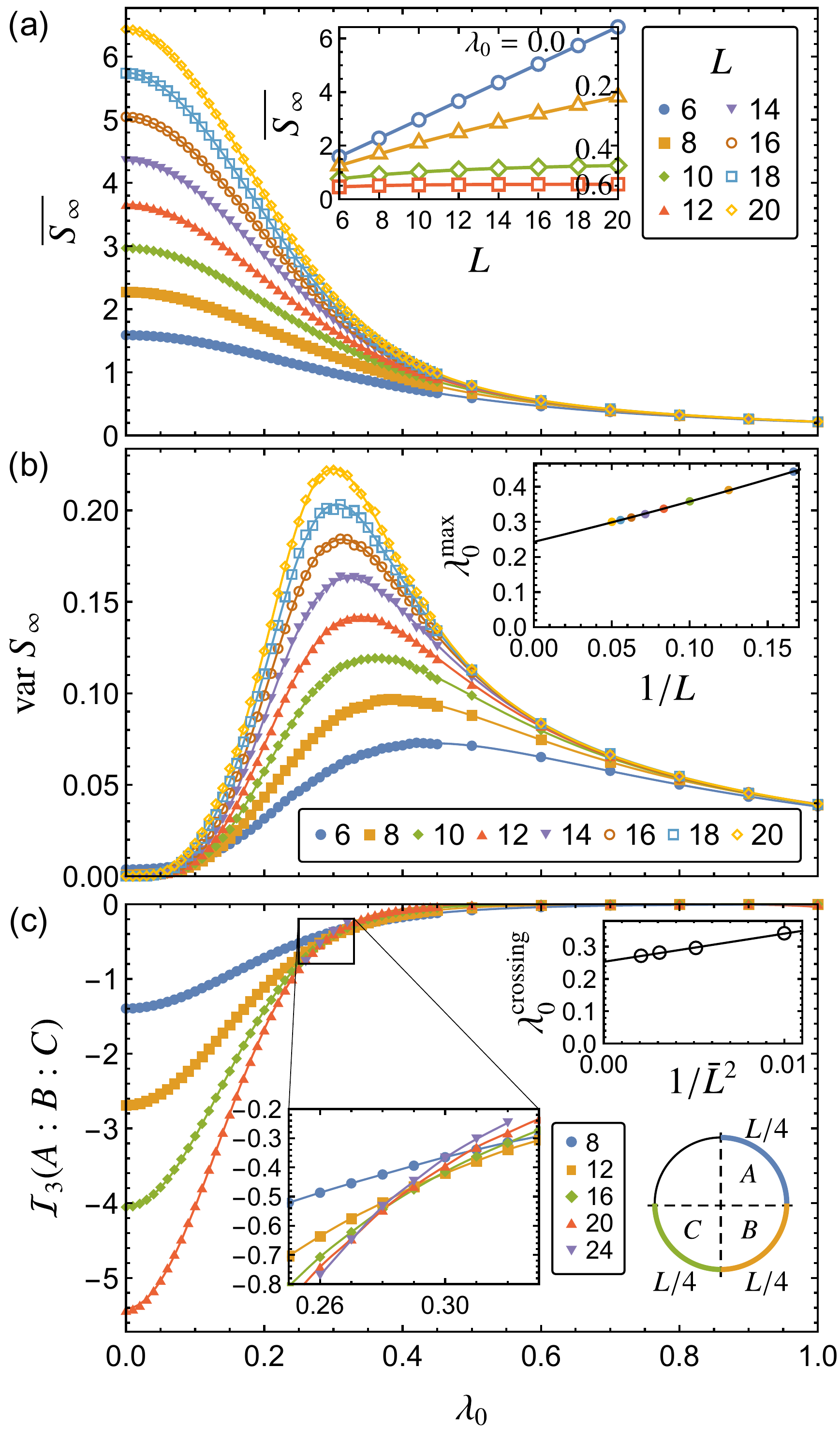}

  \caption{(a) Average saturation entropy $\overline{S_\infty}$ and (b)
  corresponding fluctuations  $\mathrm{var}\, S_\infty$, as a function of
  measurement strength $\lambda_0$ for different system sizes $L$. The inset in
  (a) shows $\overline{S_\infty}$ for fixed $\lambda_0$ as $L$ is increased,
  whilst the inset (b) shows the extrapolation of the position
  $\lambda_0^{\max}$ of maximal variance to an infinite system size. (c)
  Tripartite mutual information $\mathcal{I}_3 (A : B : C)$  as a function of
  measurement strength $\lambda_0$ for different system sizes $L$, where the
  subsystems are all of  size $L / 4$, as indicated in the bottom right inset. The left inset focuses on the region
  where the curves cross, while the top right inset shows the extrapolation of the
  crossing positions to an infinite system size, where $\bar{L}$ is the average of
  the system sizes for which the crossing occurs. \label{fig:entropy}}
\end{figure}

\emph{Entanglement dynamics in the continuum limit}.---Figure~\ref{fig:SvsTime}
illustrates the effect of the described modifications on the entanglement
dynamics in terms of the entanglement entropy for a subsystem of size $L/2$,
averaged over 1000 realizations of the dynamics initialized to a N{\'e}el state.
Time is measured as $t=n\mu^2$; in each panel, $\lambda_0=\lambda/\mu$ is kept
fixed, whilst each curve corresponds to a different value of $\mu$.

In all cases, the entanglement entropy initially increases with time, but then
saturates in a quasi-stationary regime, at a measurement-strength dependent
value $S_\infty$ that we will analyze further  in the pursuit of the
entanglement transition. For the moment, the key point in the figure is the
collapse of curves at \red{$\mu\lesssim 0.1$}, which therefore delineates
the continuum regime. That this collapse occurs both for the rate of
entanglement spreading as well as for the saturation value confirms the
described scaling of time and measurement strength in this regime. Outside of
the continuum regime, the entanglement dynamics display a notable dependence on
$\mu$, both in the rate of initial entanglement spreading as well as for the
saturation value, aspects to which we return later when we discuss the
connection to the stroboscopic case. First, we establish that an entanglement
transition occurs in the continuum regime, for which we set $\mu=0.05$
(equivalently, $dt=0.0025$).

This is demonstrated in Fig.~\ref{fig:entropy}, which shows the average and
variance of the saturation value $S_\infty$ of the entanglement entropy for
different system sizes $L$ as a function of the measurement strength
$\lambda_0$. As seen in panel (a), the entanglement entropy is large and
increases with system size when the measurement strength is small, but drops to
a small, system-size-independent value when the measurement strength is large.
As further illustrated in the inset, this qualitative change of the scaling
occurs in the range $0.2<\lambda_0<0.4$. Panel (b) shows that the
sample-to-sample
fluctuations $\mathrm{var}\,S_\infty$ indeed become large in this range.  Whilst
the position $\lambda_0^\mathrm{max}$ where the  fluctuations are maximal drifts
to smaller values as $L$ is increased, its extrapolation to infinite system size
(inset) yields a finite critical value $\lambda_0^{\text{crit}} \approx 0.243
(4)$. \red{Using this critical value for finite-size scaling
%
yields the critical exponent of the correlation length
$\nu=0.70(1)$  \cite{Note2}.} Panel (c) provides further evidence for the transition in terms
of the tripartite mutual information
\begin{eqnarray}
  \mathcal{I}_3 (A : B : C) & = &  S (A) + S (B) + S (C)+  S (A \cup B
  \cup C)\nonumber
  \\
   & - & S (A \cup B) - S (A \cup C) - S (B \cup C),
\end{eqnarray}
defined for three such subsystems $A$, $B$, and $C$. Here, the transition is
indicated by the crossing point, which has been found to show reduced
finite-size effects in the original stroboscopic model \cite{Zabalo2020}. As
shown in the insets, these features also hold in the present model, with the
position of the crossings approaching a critical value of
\red{$\lambda_0^\text{crit} =
0.253(2)$} that agrees well with the transition point obtained using the
variance analysis
\footnote{See the Supplemental Material for an analysis of
the transition in terms of the bipartite information, a discussion of
the role of measurement frequency, \red{the description and further discussion
of the finite-size scaling, and a comparison between different stochastic
processes that can be used to approach the continuum limit.
This Supplemental Material also includes a reference to Ref.
\cite{Haake1996}.}}.

\begin{figure}[t]
  \centering
  \includegraphics[width=\linewidth]{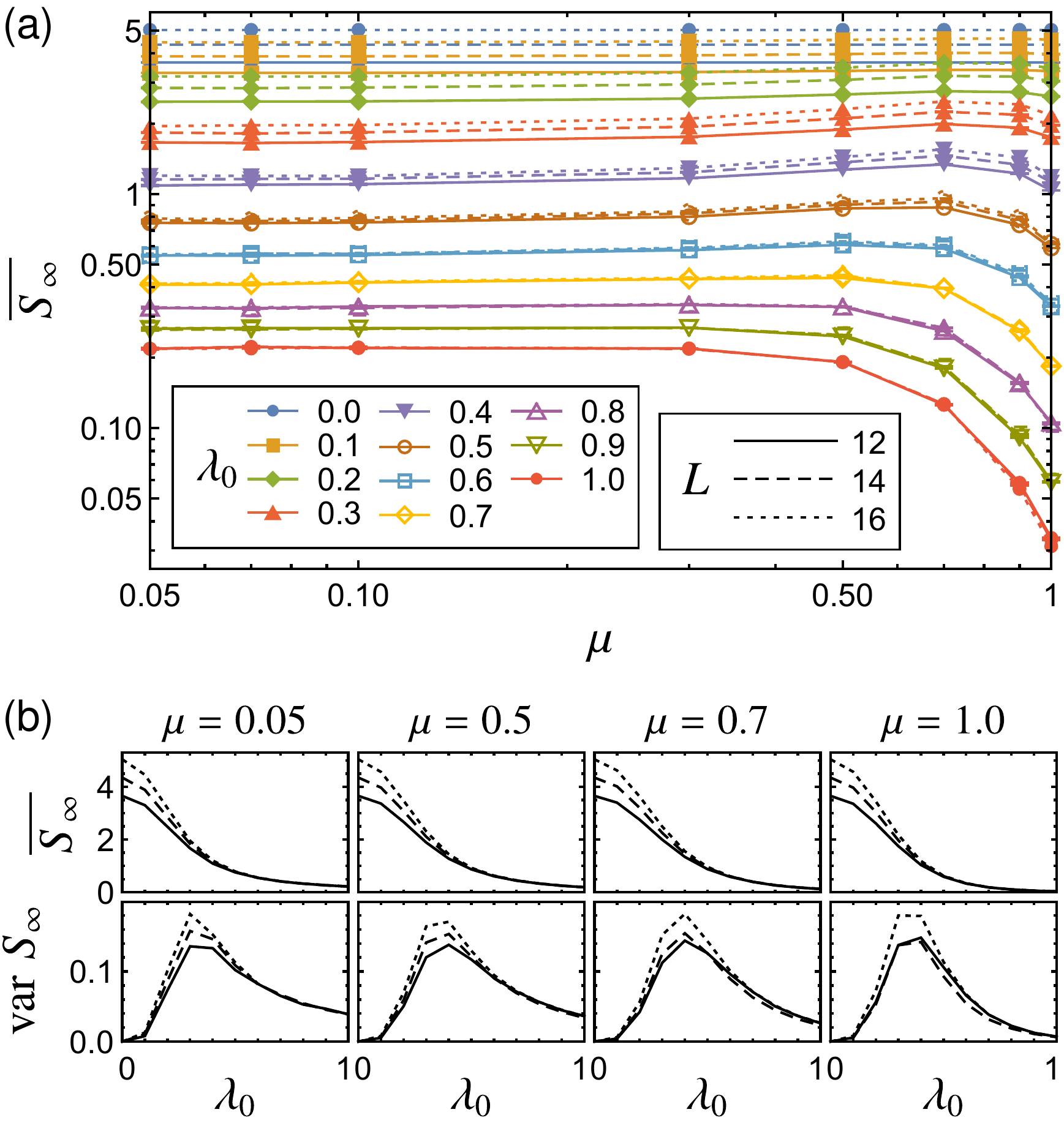}

  \caption{(a) Averaged saturation value $\overline{S_\infty}$ of the
  entanglement entropy  for different values of $\mu$ in the whole range from
  the continuum limit ($\mu\to 0)$ to the fully random stroboscopic case
  $(\mu=1$). In each curve, the measurement strength $\lambda_0=\lambda/\mu$ and
  the system size are kept fixed. Across the whole range of $\mu$, the
  entanglement entropy changes its scaling from extensive to subextensive around
  $\lambda_0\approx 0.3$. As shown in the subpanels in (b), the qualitative
  entanglement characteristics in the continuum regime
  [Fig.~\ref{fig:entropy}(a) and (b)] indeed also occur for intermediate values
  ($\mu=0.5$ and $\mu=0.7$),  with $\mu=1$ reproducing the conventional
  stroboscopic case.\label{fig:SvsT}}
\end{figure}

\emph{Connection to the stroboscopic case}.---Having established the
entanglement transition in the continuum regime, we now come to the second main
point of this paper, namely, its connection to the transition in the original
stroboscopic model. This is afforded in our model by being able to tune the
timescale $dt=\mu^2$ from $0$ to $1$. Returning to Fig.~\ref{fig:SvsTime},
outside the continuum regime the measurements still have the effect to suppress
the saturation entropy, but down to even smaller, $\mu$-dependent, values. For a
detailed analysis, Figure~\ref{fig:SvsT} (a) shows how the saturation
entanglement entropy changes with $\mu$ for fixed $\lambda_0$, where
differently-articulated curves correspond to different  system sizes $L$.
Depending on the measurement strength, we find two scenarios. For
$\lambda_0\lesssim 0.4$, the saturation entropy remains essentially
$\mu$-independent, and shows a systematic system-size dependence with an
extensive scaling, corresponding to ergodic behavior. For larger measurement
strengths, on the other hand, the entropy displays the above-mentioned downturn
as one approaches the stroboscopic limit---but also becomes independent of the
system size across the whole parameter range. As shown in the subpanels in (b),
in the intermediate range between the continuum regime and the stroboscopic
case, the average and fluctuations of the entanglement entropy display the same
qualitative behavior as in Fig.~\ref{fig:entropy}, with an only weak $\mu$
dependence of the critical value $\lambda_0^\mathrm{crit}$. These results
demonstrate a substantial degree of universality of the entanglement transition
in the whole range from the continuum regime to the fully random stroboscopic
case.

\emph{Conclusions.}---In summary, we showed that measurement-driven entanglement
transitions can occur in continuously evolving and monitored systems. We
established this in a flexible extension of random-circuit models, by which we
could directly related the transition to the widely studied stroboscopic case.
This uncovered a significant degree of university in the entanglement dynamics.
As we show in Ref.~\cite{Note2}, this universality further extends to the
variation of the measurement frequency $p$ (the parameter that was varied in the
original studies of the stroboscopic model), where results remain invariant upon
a simple rescaling $\lambda_0=\sqrt{p}\lambda/\mu$ of the effective measurement
strength. In this way, results derived for stroboscopic models gain a much
larger range of applicability.

The model described in this work has been designed to not only interpolate
between different scenarios, but also to combine the most  generic effects of
random dynamics and continuous measurements, and thereby, to further inform the
design of suitable experiments. In particular, the unitary dynamics describe the
local generation of entanglement by randomly fluctuating interactions, whilst
the employed measurement model describes quantum detection schemes currently
employed in solid state nanocircuits \cite{Field1993measurements,
Korotkov1999continuous, Romito2008weak, Murch2013observing, Aasen2016milestones}
and quantum optical devices \cite{Hosten2008observation,
Dixon2009ultrasensitive}. In such settings, the described universality of the
entanglement dynamics enhances our understanding of environmental effects, and
serves to provide detailed control of the quantum dynamics in simple yet
profound ways.

\begin{acknowledgments}
This research was funded by the UK Engineering and Physical Sciences Research
Council (EPSRC) via Grant No.~EP/P010180/1. Computer time was provided by
Lancaster University's High-End Computing facility. M.S. was also funded by the 
European Research Council (ERC) under the European Union's Horizon 2020 
Research and Innovation programme (Grant agreement No. 853368). All relevant 
data present in this publication can be accessed at 
\url{https://dx.doi.org/10.17635/lancaster/researchdata/396}.
\end{acknowledgments}

\section{Appendix A: Further entanglement characteristics}

\begin{figure}[b]
  \centering
  \includegraphics[width=0.8\linewidth]{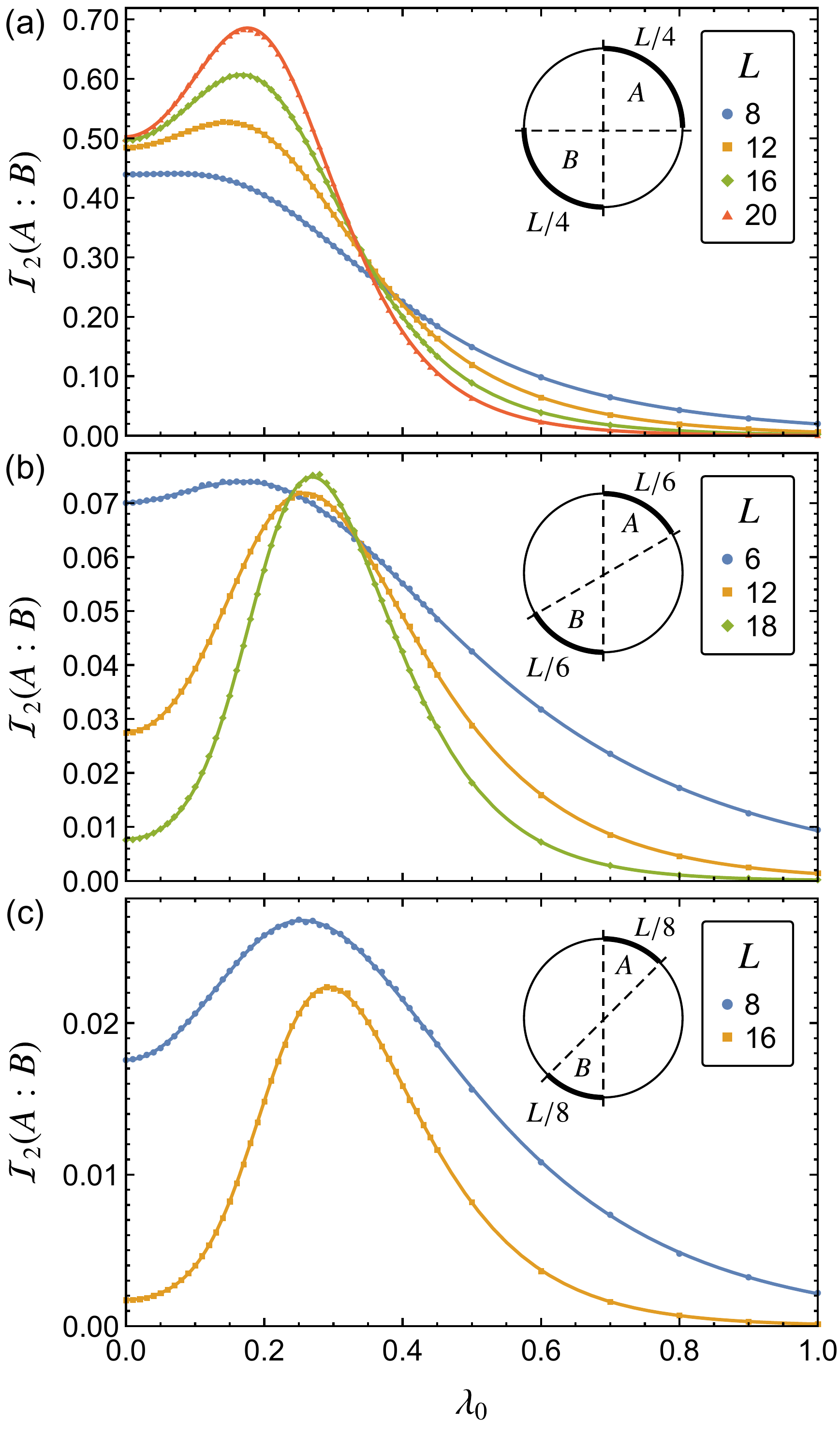}

  \caption{Bipartite mutual information $\mathcal{I}_2 (A : B)$ for antipodal
  subsystems of size (a) $L / 4$, (b) $L / 6$, and (c) $L / 8$.\label{fig:mi2}}
\end{figure}

Figure \ref{fig:mi2} presents our data for the bipartite mutual information
\begin{equation}
  \mathcal{I}_2 (A : B) = S (A) + S (B) - S (A \cup B).
\end{equation}
This is defined for two  subsystems $A$ and $B$, here chosen to be antipodal,
whose size is varied proportionally to the overall system size in the ratio of
(a) $1 / 4$, (b) $1 / 6$, and (c) $1 / 8$. Whilst each case provides
characteristic maxima and crossings in the range $0.2\lesssim \lambda_0 \lesssim
0.4$, these occur at different locations and show inconsistent finite-size
effects, making this quantity less suitable to determine the transition when
compared to the bipartite entropy and tripartite mutual information studied in
the main text.

\changes{We also note that the Hartley entropy [$(n=0)$-R\'enyi entropy], which
for projective measurements exhibits a phase transition at a different 
critical frequency $p_c$ than other R\'enyi entropies~\cite{Skinner2018, 
Zabalo2020}, in our model is always maximal for any $\lambda_0<\infty$, as the 
weak measurements never completely break bonds in the circuit.}

\section{Appendix B:  Measurement frequency}

\begin{figure}[t]
  \centering
  \includegraphics[width=0.7\linewidth]{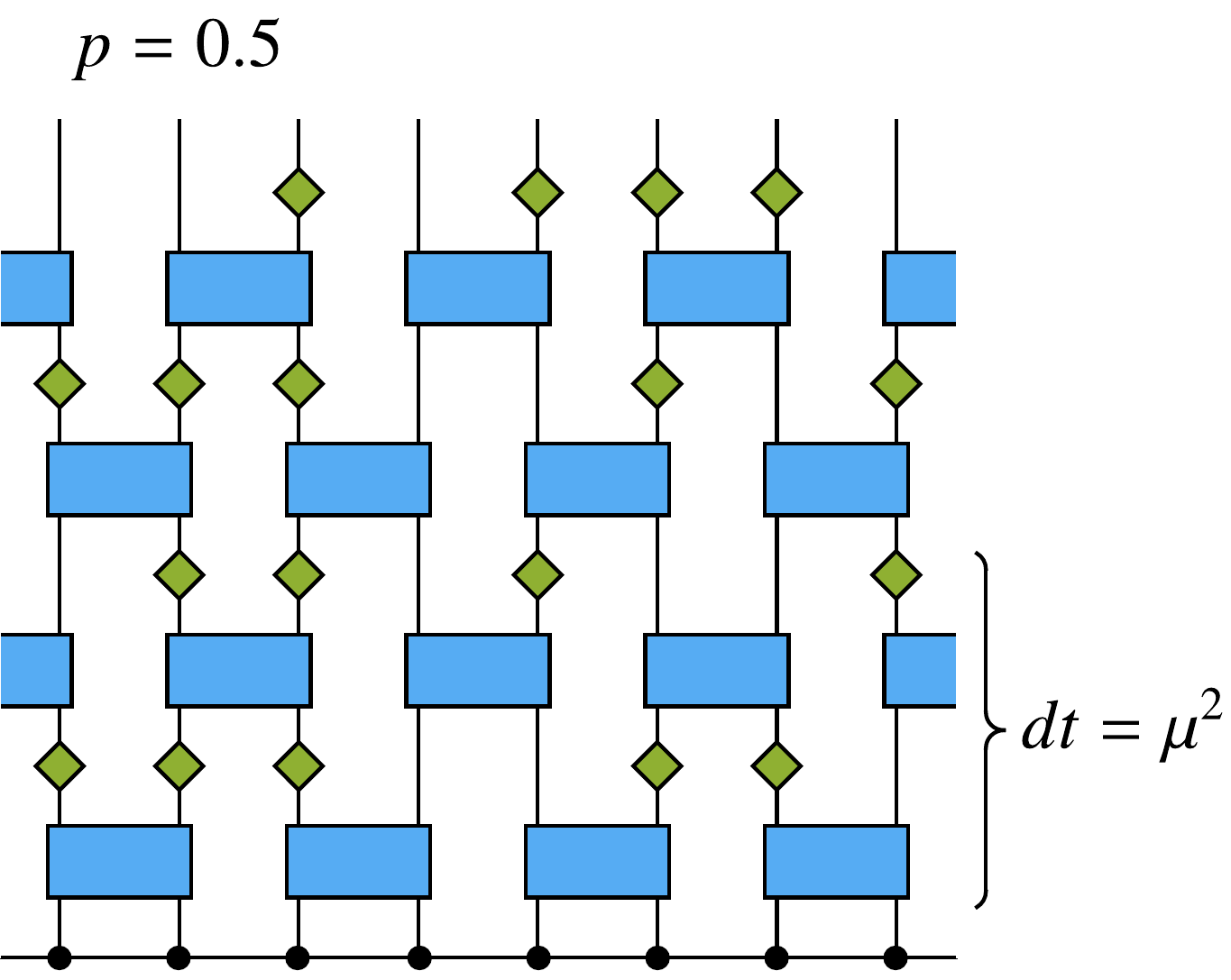}

  \caption{Illustration of a random circuit with measurement frequency $p = 1 /
  2$. In the continuum limit, this results in the same behavior as the original
  circuit  with altered effective measurement strength $ \lambda_0^{\text{eff}} 
  =\sqrt{p}\lambda/\mu$.
  \label{fig:frequency_circuit}}
\end{figure}

In this Appendix, we address the issue of applying the measurements $M$ with a
specific frequency $p \in [0, 1]$, a question that has been explored in many
recent studies in the original stroboscopic model~{\cite{Li2018, Chan2018,
Skinner2018, Li2019, Szyniszewski2019}}. For illustration, a modified circuit
with $p = 1 / 2$ is shown in Fig.~\ref{fig:frequency_circuit}.

\begin{figure}[t]
  \raggedright
  \includegraphics[width=0.9\linewidth]{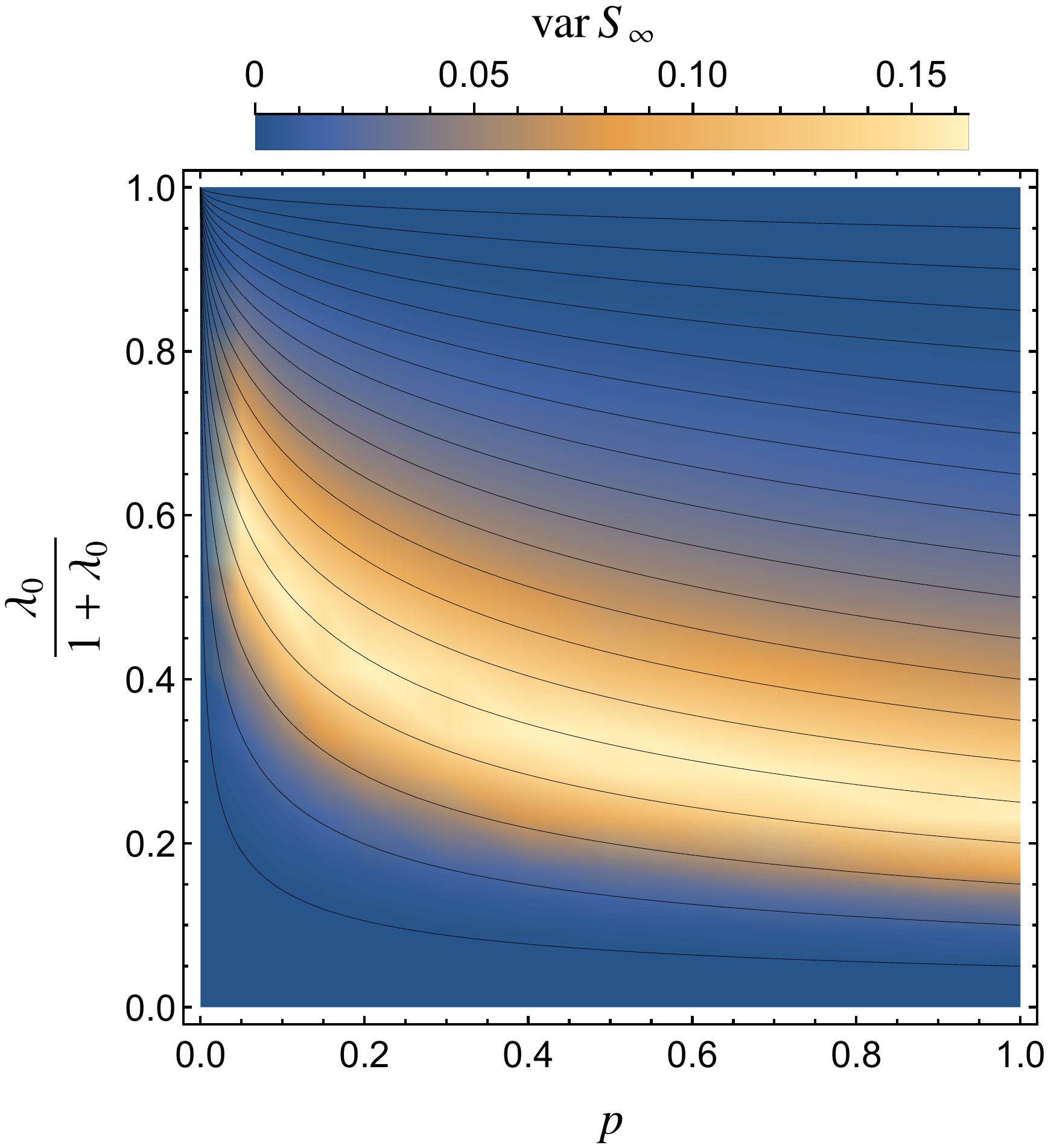}
  \caption{Contour plot of the variance $\mathrm{var}\,S_\infty$ of the
  saturation entropy as a function of
  measurement frequency $p$ and the (not rescaled) measurement strength
  $\lambda_0=\lambda/\mu$, for
   system size $L = 14$. The black lines show contours of equivalent
  systems with fixed $ \lambda_0^{\text{eff}} =\sqrt{p}\lambda_0$.
  \label{fig:p_vs_lambda0}}
\end{figure}

\begin{figure*}[t]
  \centering
  \includegraphics[width=\linewidth]{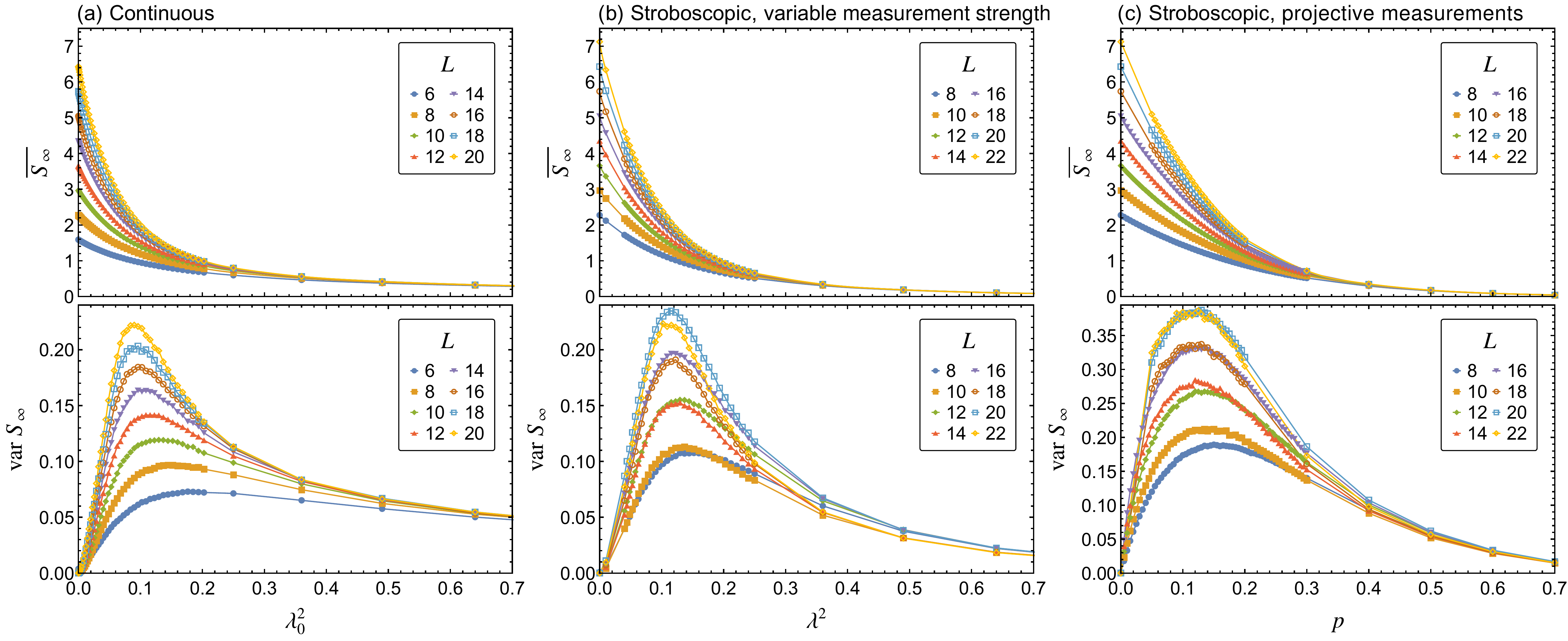}

  \caption{Comparison of the entanglement characteristics from (a) the model of
  the main text in the continuum limit ($\mu,\lambda\ll 1$  at fixed
  $\lambda_0=\lambda/\mu$) as well as the stroboscopic case ($\mu=1$) with (b)
  measurements of variable strength $\lambda$ or (c) projective measurements
  with rate $p$. \label{fig:lambda_squared}}
\end{figure*}

As shown in Fig.~\ref{fig:p_vs_lambda0}, the entanglement characteristics of the
system remain invariant for a constant value of
\begin{equation}
  \lambda_0^{\text{eff}} = \sqrt{p}\lambda/\mu,
  \label{eq:strengthp}
\end{equation}
which becomes exact in the continuum limit, and is this key feature mentioned in
the conclusions. For completeness, we show in Fig.~\ref{fig:lambda_squared} that
the entanglement characteristics from our model in the continuum limit, panel
(a), are qualitatively the same as obtained in the stroboscopic case with
variable-strength measurements (b) or projective measurements with rate $p$ (c),
where parameters are rescaled  in analogy to Eq.~\eqref{eq:strengthp}.

\red{
\section{Appendix C:  Extrapolation to the infinite system size}

This Appendix details the fitting procedures used to extract the
infinite-system-size values in Fig.~3. Both $\lambda_0^{\rm max}$ and
$\lambda_0^{\rm crossing}$ have an unknown dependence on the system
size. We, therefore, used a standard polynomial fitting procedure in
$1/L$.

By analyzing the fits for $\lambda_0^{\rm max}$ with polynomial forms
up to the first, the second and the third order in $1/L$, we have
determined that the linear fit underfits the data, while the fit with
terms up to $1/L^3$ overfits the data. The quadratic fit was found to
match the data best. We note that the polynomial expansion of
$\lambda_0^{\rm max}$ in $1/L$ is also motivated by previous numerical
results in Ref.~\cite{Szyniszewski2019} (circuit with stroboscopic
weak measurements), where $\lambda^{\rm max}$ was found to be a
polynomial function of $1/L$, while $p^{\rm max}$ -- a polynomial
function of $1/L^2$. This is consistent with our discussions in
Appendix B, where we show that $\lambda_0$ from the continuum limit
behaves like $\lambda$ in the stroboscopic limit with $p=1$,
and furthermore scales with $\sqrt{p}$ in the stroboscopic limit with 
$\lambda\to\infty$.

A similar analysis for $\lambda_0^{\rm crossing}$ reveals that the
linear contribution in $1/L$ is statistically irrelevant. We therefore
adopted a polynomial fit in $1/L^2$, and after similar
investigation as above, a linear fit was found to describe the data
best, while quadratic fit in $1/L^2$ was overfitting our data.

\section{Appendix D:  Finite size scaling of the entanglement entropy}

In order to extract the critical exponent of $\nu$ of the correlation
length, we attempted finite size scaling of the entanglement entropy,
using the following scaling ansatz proposed in Ref.~\cite{Skinner2018}:
\begin{equation}
  | \overline{S_\infty}(\lambda_0)-\overline{S_\infty}(\lambda_0^{\rm crit}) |
  = F\left[ (\lambda_0-\lambda_0^{\rm crit}) L^{1/\nu} \right],
  \label{eq:fss}
\end{equation}
where $F$ is an unknown \changes{one-parameter} scaling function. Assuming
$\lambda_0^{\rm crit} = 0.243$ extracted from the variance analysis (see the
main text), the data collapse [shown in Fig.~\ref{fig:fss}(a)] gives
$\nu=0.70(1)$. \changes{The same scaling for the stroboscopic
limit~\cite{Szyniszewski2019} with $\lambda_0 = 0.301$ yields $\nu=0.80(3)$ [see
Fig.~\ref{fig:fss}(b)]. The two estimates of $\nu$ are close, and since we do
not account for all finite size errors, it may be that both critical exponents
agree.}

\begin{figure}[t]
  \raggedright
  \includegraphics[width=0.9\linewidth]{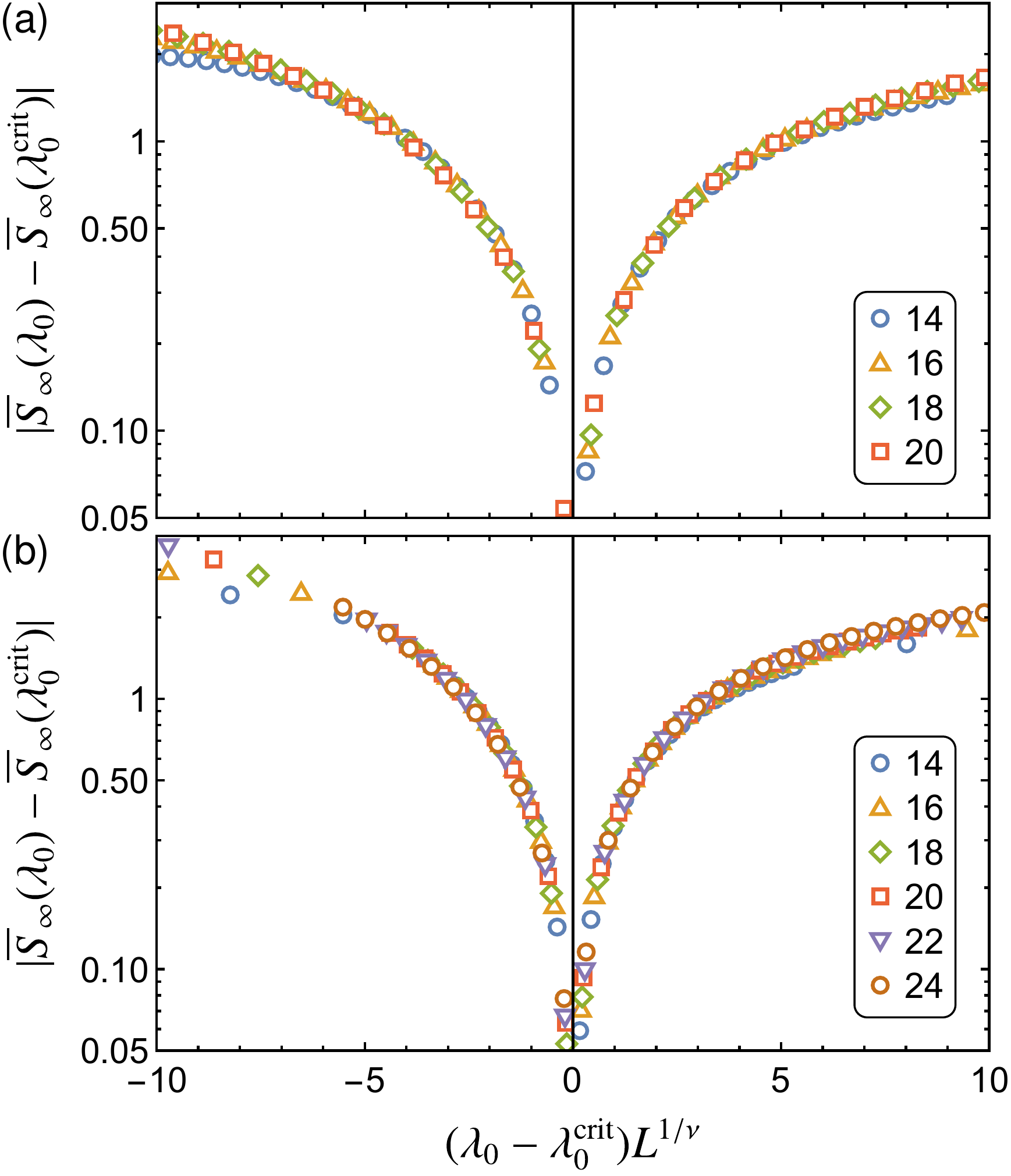}
  \caption{
  Finite size scaling of the saturation entanglement entropy \changes{using the 
  scaling from Eq.~(\ref{eq:fss}) for (a) the continuum limit ($\mu\to 0$) and 
  (b) the stroboscopic model ($\mu=1$).}
  \label{fig:fss}}
\end{figure}

\section{Appendix E: Comparison between the Cauchy process and Wiener process}

In the main text, we approach the limit of continuous stochastic quantum 
dynamics based 
on a Cauchy process, which allows us to interpolate between this limit and the 
stroboscopic case with random unitaries distributed by the Haar measure. In 
order to assert that the results from  this process are generic, we compare 
these in this Appendix to those of a Wiener process.
This Wiener process is realized by replacing $H_\mathrm{eff}$ in
Eq.~(3) by an $N\times N$-dimensional Hermitian matrix with 
Gaussian-distributed elements with $\langle 
|H_{\mathrm{eff},lm}|^2\rangle=\sigma^2/dt$, where $N=4$ for dynamics generated 
by acting onto two spins.
Therefore, in the Wiener process matrix elements have to be scaled as
$H_{\mathrm{eff},lm}\sim \mu^{-1}$, which is as expected given that 
$\mu^{-1}=dt^{-1/2}$.
Indeed, the processes can be related by comparing the ensemble averages of the 
resulting unitary matrices, for which we find
\begin{align}
\langle |U_{nn}|^2\rangle&=
\left\{
  \begin{array}{ll}
    1-dt(N-1)\sigma^2  & \mbox{(Wiener);} \\
    1-dt(N-1)/(N+1) & \mbox{(Cauchy);}
  \end{array}
\right.
\\
\langle |U_{n\neq m}|^2\rangle&=
\left\{
  \begin{array}{ll}
    dt\,\sigma^2  & \mbox{(Wiener);} \\
    dt/(N+1)& \mbox{(Cauchy).}
  \end{array}
\right.
\end{align}
(for the Cauchy process we adopted results from \cite{Haake1996}). 

We confirmed this scaling numerically by matching the transient dynamical 
entanglement growth rate in both processes, which we found to agree with a 
fixed proportionality constant that is independent of the overall system size 
and other parameters of the model. As shown in Fig.~\ref{fig:wiener}, using 
this input from the transient dynamics, we furthermore find exactly the same 
dependence of the averaged entropy and mutual information in both processes. 
Finally, the fluctuations of the entropy, which further probe details of the 
temporal correlations, match when rescaled by a multiplicative factor of order 
of unity, which is again the same for all parameters and system sizes.
These observations further support the universal character of the entanglement 
transition, which displays the same generic behavior across all described 
scenarios.

\begin{figure}[t]
  \includegraphics[width=\linewidth]{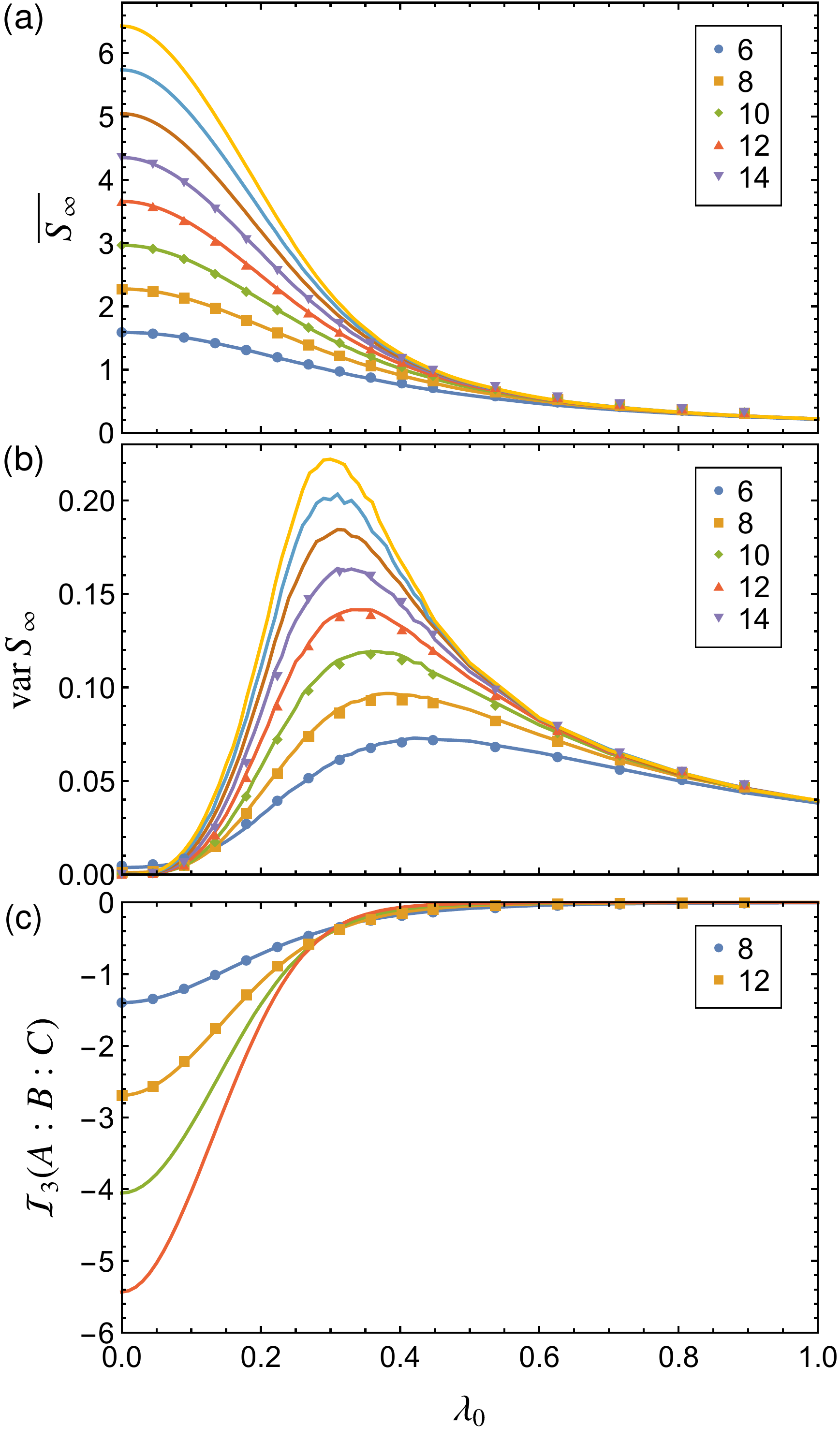}
  \caption{
  Comparison between the Cauchy process (lines) and the Wiener process
  (markers) for (a) average saturation entropy
  $\overline{S_{\infty}}$, (b) corresponding fluctuations $\text{var}
  S_{\infty}$, and (c) tripartite mutual information $\mathcal{I}_3 (A
  : B : C)$. The size of the fluctuations for the Cauchy process was
  normalized to match the Wiener process.\label{fig:wiener}}
\end{figure}
}

\bibliography{refs}

\begin{thebibliography}{74}%
\makeatletter
\providecommand \@ifxundefined [1]{%
 \@ifx{#1\undefined}
}%
\providecommand \@ifnum [1]{%
 \ifnum #1\expandafter \@firstoftwo
 \else \expandafter \@secondoftwo
 \fi
}%
\providecommand \@ifx [1]{%
 \ifx #1\expandafter \@firstoftwo
 \else \expandafter \@secondoftwo
 \fi
}%
\providecommand \natexlab [1]{#1}%
\providecommand \enquote  [1]{``#1''}%
\providecommand \bibnamefont  [1]{#1}%
\providecommand \bibfnamefont [1]{#1}%
\providecommand \citenamefont [1]{#1}%
\providecommand \href@noop [0]{\@secondoftwo}%
\providecommand \href [0]{\begingroup \@sanitize@url \@href}%
\providecommand \@href[1]{\@@startlink{#1}\@@href}%
\providecommand \@@href[1]{\endgroup#1\@@endlink}%
\providecommand \@sanitize@url [0]{\catcode `\\12\catcode `\$12\catcode
  `\&12\catcode `\#12\catcode `\^12\catcode `\_12\catcode `\%12\relax}%
\providecommand \@@startlink[1]{}%
\providecommand \@@endlink[0]{}%
\providecommand \url  [0]{\begingroup\@sanitize@url \@url }%
\providecommand \@url [1]{\endgroup\@href {#1}{\urlprefix }}%
\providecommand \urlprefix  [0]{URL }%
\providecommand \Eprint [0]{\href }%
\providecommand \doibase [0]{https://doi.org/}%
\providecommand \selectlanguage [0]{\@gobble}%
\providecommand \bibinfo  [0]{\@secondoftwo}%
\providecommand \bibfield  [0]{\@secondoftwo}%
\providecommand \translation [1]{[#1]}%
\providecommand \BibitemOpen [0]{}%
\providecommand \bibitemStop [0]{}%
\providecommand \bibitemNoStop [0]{.\EOS\space}%
\providecommand \EOS [0]{\spacefactor3000\relax}%
\providecommand \BibitemShut  [1]{\csname bibitem#1\endcsname}%
\let\auto@bib@innerbib\@empty
\bibitem [{\citenamefont {Degasperis}\ \emph {et~al.}(1974)\citenamefont
  {Degasperis}, \citenamefont {Fonda},\ and\ \citenamefont
  {Ghirardi}}]{Degasperis1974}%
  \BibitemOpen
  \bibfield  {author} {\bibinfo {author} {\bibfnamefont {A.}~\bibnamefont
  {Degasperis}}, \bibinfo {author} {\bibfnamefont {L.}~\bibnamefont {Fonda}},\
  and\ \bibinfo {author} {\bibfnamefont {G.~C.}\ \bibnamefont {Ghirardi}},\
  }\bibfield  {title} {\bibinfo {title} {Does the lifetime of an unstable
  system depend on the measuring apparatus?},\ }\href
  {https://doi.org/10.1007/BF02731351} {\bibfield  {journal} {\bibinfo
  {journal} {Nuov. Cim. A}\ }\textbf {\bibinfo {volume} {21}},\ \bibinfo
  {pages} {471} (\bibinfo {year} {1974})}\BibitemShut {NoStop}%
\bibitem [{\citenamefont {Misra}\ and\ \citenamefont
  {Sudarshan}(1977)}]{Misra1977}%
  \BibitemOpen
  \bibfield  {author} {\bibinfo {author} {\bibfnamefont {B.}~\bibnamefont
  {Misra}}\ and\ \bibinfo {author} {\bibfnamefont {E.~C.~G.}\ \bibnamefont
  {Sudarshan}},\ }\bibfield  {title} {\bibinfo {title} {{The Zeno's paradox in
  quantum theory}},\ }\href {https://doi.org/10.1063/1.523304} {\bibfield
  {journal} {\bibinfo  {journal} {J. Math. Phys.}\ }\textbf {\bibinfo {volume}
  {18}},\ \bibinfo {pages} {756} (\bibinfo {year} {1977})}\BibitemShut
  {NoStop}%
\bibitem [{\citenamefont {Peres}(1980)}]{Peres1980}%
  \BibitemOpen
  \bibfield  {author} {\bibinfo {author} {\bibfnamefont {A.}~\bibnamefont
  {Peres}},\ }\bibfield  {title} {\bibinfo {title} {Zeno paradox in quantum
  theory},\ }\href {https://doi.org/10.1119/1.12204} {\bibfield  {journal}
  {\bibinfo  {journal} {Am. J. Phys.}\ }\textbf {\bibinfo {volume} {48}},\
  \bibinfo {pages} {931} (\bibinfo {year} {1980})}\BibitemShut {NoStop}%
\bibitem [{\citenamefont {Li}\ \emph {et~al.}(2018)\citenamefont {Li},
  \citenamefont {Chen},\ and\ \citenamefont {Fisher}}]{Li2018}%
  \BibitemOpen
  \bibfield  {author} {\bibinfo {author} {\bibfnamefont {Y.}~\bibnamefont
  {Li}}, \bibinfo {author} {\bibfnamefont {X.}~\bibnamefont {Chen}},\ and\
  \bibinfo {author} {\bibfnamefont {M.~P.~A.}\ \bibnamefont {Fisher}},\
  }\bibfield  {title} {\bibinfo {title} {{Quantum Zeno effect and the many-body
  entanglement transition}},\ }\href
  {https://doi.org/10.1103/PhysRevB.98.205136} {\bibfield  {journal} {\bibinfo
  {journal} {Phys. Rev. B}\ }\textbf {\bibinfo {volume} {98}},\ \bibinfo
  {pages} {205136} (\bibinfo {year} {2018})}\BibitemShut {NoStop}%
\bibitem [{\citenamefont {Chan}\ \emph {et~al.}(2019)\citenamefont {Chan},
  \citenamefont {Nandkishore}, \citenamefont {Pretko},\ and\ \citenamefont
  {Smith}}]{Chan2018}%
  \BibitemOpen
  \bibfield  {author} {\bibinfo {author} {\bibfnamefont {A.}~\bibnamefont
  {Chan}}, \bibinfo {author} {\bibfnamefont {R.~M.}\ \bibnamefont
  {Nandkishore}}, \bibinfo {author} {\bibfnamefont {M.}~\bibnamefont
  {Pretko}},\ and\ \bibinfo {author} {\bibfnamefont {G.}~\bibnamefont
  {Smith}},\ }\bibfield  {title} {\bibinfo {title} {Unitary-projective
  entanglement dynamics},\ }\href {https://doi.org/10.1103/PhysRevB.99.224307}
  {\bibfield  {journal} {\bibinfo  {journal} {Phys. Rev. B}\ }\textbf {\bibinfo
  {volume} {99}},\ \bibinfo {pages} {224307} (\bibinfo {year}
  {2019})}\BibitemShut {NoStop}%
\bibitem [{\citenamefont {Skinner}\ \emph {et~al.}(2019)\citenamefont
  {Skinner}, \citenamefont {Ruhman},\ and\ \citenamefont
  {Nahum}}]{Skinner2018}%
  \BibitemOpen
  \bibfield  {author} {\bibinfo {author} {\bibfnamefont {B.}~\bibnamefont
  {Skinner}}, \bibinfo {author} {\bibfnamefont {J.}~\bibnamefont {Ruhman}},\
  and\ \bibinfo {author} {\bibfnamefont {A.}~\bibnamefont {Nahum}},\ }\bibfield
   {title} {\bibinfo {title} {Measurement-induced phase transitions in the
  dynamics of entanglement},\ }\href
  {https://doi.org/10.1103/PhysRevX.9.031009} {\bibfield  {journal} {\bibinfo
  {journal} {Phys. Rev. X}\ }\textbf {\bibinfo {volume} {9}},\ \bibinfo {pages}
  {031009} (\bibinfo {year} {2019})}\BibitemShut {NoStop}%
\bibitem [{\citenamefont {Li}\ \emph {et~al.}(2019)\citenamefont {Li},
  \citenamefont {Chen},\ and\ \citenamefont {Fisher}}]{Li2019}%
  \BibitemOpen
  \bibfield  {author} {\bibinfo {author} {\bibfnamefont {Y.}~\bibnamefont
  {Li}}, \bibinfo {author} {\bibfnamefont {X.}~\bibnamefont {Chen}},\ and\
  \bibinfo {author} {\bibfnamefont {M.~P.~A.}\ \bibnamefont {Fisher}},\
  }\bibfield  {title} {\bibinfo {title} {Measurement-driven entanglement
  transition in hybrid quantum circuits},\ }\href
  {https://doi.org/10.1103/PhysRevB.100.134306} {\bibfield  {journal} {\bibinfo
   {journal} {Phys. Rev. B}\ }\textbf {\bibinfo {volume} {100}},\ \bibinfo
  {pages} {134306} (\bibinfo {year} {2019})}\BibitemShut {NoStop}%
\bibitem [{\citenamefont {Szyniszewski}\ \emph {et~al.}(2019)\citenamefont
  {Szyniszewski}, \citenamefont {Romito},\ and\ \citenamefont
  {Schomerus}}]{Szyniszewski2019}%
  \BibitemOpen
  \bibfield  {author} {\bibinfo {author} {\bibfnamefont {M.}~\bibnamefont
  {Szyniszewski}}, \bibinfo {author} {\bibfnamefont {A.}~\bibnamefont
  {Romito}},\ and\ \bibinfo {author} {\bibfnamefont {H.}~\bibnamefont
  {Schomerus}},\ }\bibfield  {title} {\bibinfo {title} {Entanglement transition
  from variable-strength weak measurements},\ }\href
  {https://doi.org/10.1103/PhysRevB.100.064204} {\bibfield  {journal} {\bibinfo
   {journal} {Phys. Rev. B}\ }\textbf {\bibinfo {volume} {100}},\ \bibinfo
  {pages} {064204} (\bibinfo {year} {2019})}\BibitemShut {NoStop}%
\bibitem [{\citenamefont {Jian}\ \emph {et~al.}(2020)\citenamefont {Jian},
  \citenamefont {You}, \citenamefont {Vasseur},\ and\ \citenamefont
  {Ludwig}}]{Jian2020}%
  \BibitemOpen
  \bibfield  {author} {\bibinfo {author} {\bibfnamefont {C.-M.}\ \bibnamefont
  {Jian}}, \bibinfo {author} {\bibfnamefont {Y.-Z.}\ \bibnamefont {You}},
  \bibinfo {author} {\bibfnamefont {R.}~\bibnamefont {Vasseur}},\ and\ \bibinfo
  {author} {\bibfnamefont {A.~W.~W.}\ \bibnamefont {Ludwig}},\ }\bibfield
  {title} {\bibinfo {title} {Measurement-induced criticality in random quantum
  circuits},\ }\href {https://doi.org/10.1103/PhysRevB.101.104302} {\bibfield
  {journal} {\bibinfo  {journal} {Phys. Rev. B}\ }\textbf {\bibinfo {volume}
  {101}},\ \bibinfo {pages} {104302} (\bibinfo {year} {2020})}\BibitemShut
  {NoStop}%
\bibitem [{\citenamefont {Li}\ \emph {et~al.}(2020)\citenamefont {Li},
  \citenamefont {Chen}, \citenamefont {Ludwig},\ and\ \citenamefont
  {Fisher}}]{Li2020}%
  \BibitemOpen
  \bibfield  {author} {\bibinfo {author} {\bibfnamefont {Y.}~\bibnamefont
  {Li}}, \bibinfo {author} {\bibfnamefont {X.}~\bibnamefont {Chen}}, \bibinfo
  {author} {\bibfnamefont {A.~W.~W.}\ \bibnamefont {Ludwig}},\ and\ \bibinfo
  {author} {\bibfnamefont {M.~P.~A.}\ \bibnamefont {Fisher}},\ }\bibfield
  {title} {\bibinfo {title} {Conformal invariance and quantum non-locality in
  hybrid quantum circuits},\ }\Eprint {https://arxiv.org/abs/2003.12721}
  {arXiv:2003.12721}  (\bibinfo {year} {2020})\BibitemShut {NoStop}%
\bibitem [{\citenamefont {Chen}\ \emph {et~al.}(2020)\citenamefont {Chen},
  \citenamefont {Li}, \citenamefont {Fisher},\ and\ \citenamefont
  {Lucas}}]{Chen2020}%
  \BibitemOpen
  \bibfield  {author} {\bibinfo {author} {\bibfnamefont {X.}~\bibnamefont
  {Chen}}, \bibinfo {author} {\bibfnamefont {Y.}~\bibnamefont {Li}}, \bibinfo
  {author} {\bibfnamefont {M.~P.~A.}\ \bibnamefont {Fisher}},\ and\ \bibinfo
  {author} {\bibfnamefont {A.}~\bibnamefont {Lucas}},\ }\bibfield  {title}
  {\bibinfo {title} {Emergent conformal symmetry in nonunitary random dynamics
  of free fermions},\ }\href {https://doi.org/10.1103/PhysRevResearch.2.033017}
  {\bibfield  {journal} {\bibinfo  {journal} {Phys. Rev. Research}\ }\textbf
  {\bibinfo {volume} {2}},\ \bibinfo {pages} {033017} (\bibinfo {year}
  {2020})}\BibitemShut {NoStop}%
\bibitem [{\citenamefont {Choi}\ \emph {et~al.}(2020)\citenamefont {Choi},
  \citenamefont {Bao}, \citenamefont {Qi},\ and\ \citenamefont
  {Altman}}]{Choi2019}%
  \BibitemOpen
  \bibfield  {author} {\bibinfo {author} {\bibfnamefont {S.}~\bibnamefont
  {Choi}}, \bibinfo {author} {\bibfnamefont {Y.}~\bibnamefont {Bao}}, \bibinfo
  {author} {\bibfnamefont {X.-L.}\ \bibnamefont {Qi}},\ and\ \bibinfo {author}
  {\bibfnamefont {E.}~\bibnamefont {Altman}},\ }\bibfield  {title} {\bibinfo
  {title} {Quantum error correction in scrambling dynamics and
  measurement-induced phase transition},\ }\href
  {https://doi.org/10.1103/PhysRevLett.125.030505} {\bibfield  {journal}
  {\bibinfo  {journal} {Phys. Rev. Lett.}\ }\textbf {\bibinfo {volume} {125}},\
  \bibinfo {pages} {030505} (\bibinfo {year} {2020})}\BibitemShut {NoStop}%
\bibitem [{\citenamefont {Gullans}\ and\ \citenamefont
  {Huse}(2020{\natexlab{a}})}]{Gullans2019purification}%
  \BibitemOpen
  \bibfield  {author} {\bibinfo {author} {\bibfnamefont {M.~J.}\ \bibnamefont
  {Gullans}}\ and\ \bibinfo {author} {\bibfnamefont {D.~A.}\ \bibnamefont
  {Huse}},\ }\bibfield  {title} {\bibinfo {title} {Dynamical purification phase
  transition induced by quantum measurements},\ }\href
  {https://doi.org/10.1103/PhysRevX.10.041020} {\bibfield  {journal} {\bibinfo
  {journal} {Phys. Rev. X}\ }\textbf {\bibinfo {volume} {10}},\ \bibinfo
  {pages} {041020} (\bibinfo {year} {2020}{\natexlab{a}})}\BibitemShut
  {NoStop}%
\bibitem [{\citenamefont {Kuo}\ \emph {et~al.}(2020)\citenamefont {Kuo},
  \citenamefont {Akhtar}, \citenamefont {Arovas},\ and\ \citenamefont
  {You}}]{Kuo2019}%
  \BibitemOpen
  \bibfield  {author} {\bibinfo {author} {\bibfnamefont {W.-T.}\ \bibnamefont
  {Kuo}}, \bibinfo {author} {\bibfnamefont {A.~A.}\ \bibnamefont {Akhtar}},
  \bibinfo {author} {\bibfnamefont {D.~P.}\ \bibnamefont {Arovas}},\ and\
  \bibinfo {author} {\bibfnamefont {Y.-Z.}\ \bibnamefont {You}},\ }\bibfield
  {title} {\bibinfo {title} {Markovian entanglement dynamics under locally
  scrambled quantum evolution},\ }\href
  {https://doi.org/10.1103/PhysRevB.101.224202} {\bibfield  {journal} {\bibinfo
   {journal} {Phys. Rev. B}\ }\textbf {\bibinfo {volume} {101}},\ \bibinfo
  {pages} {224202} (\bibinfo {year} {2020})}\BibitemShut {NoStop}%
\bibitem [{\citenamefont {Nahum}\ and\ \citenamefont
  {Skinner}(2020)}]{Nahum2019}%
  \BibitemOpen
  \bibfield  {author} {\bibinfo {author} {\bibfnamefont {A.}~\bibnamefont
  {Nahum}}\ and\ \bibinfo {author} {\bibfnamefont {B.}~\bibnamefont
  {Skinner}},\ }\bibfield  {title} {\bibinfo {title} {Entanglement and dynamics
  of diffusion-annihilation processes with majorana defects},\ }\href
  {https://doi.org/10.1103/PhysRevResearch.2.023288} {\bibfield  {journal}
  {\bibinfo  {journal} {Phys. Rev. Research}\ }\textbf {\bibinfo {volume}
  {2}},\ \bibinfo {pages} {023288} (\bibinfo {year} {2020})}\BibitemShut
  {NoStop}%
\bibitem [{\citenamefont {Roy}\ \emph {et~al.}(2020)\citenamefont {Roy},
  \citenamefont {Chalker}, \citenamefont {Gornyi},\ and\ \citenamefont
  {Gefen}}]{Roy2019}%
  \BibitemOpen
  \bibfield  {author} {\bibinfo {author} {\bibfnamefont {S.}~\bibnamefont
  {Roy}}, \bibinfo {author} {\bibfnamefont {J.~T.}\ \bibnamefont {Chalker}},
  \bibinfo {author} {\bibfnamefont {I.~V.}\ \bibnamefont {Gornyi}},\ and\
  \bibinfo {author} {\bibfnamefont {Y.}~\bibnamefont {Gefen}},\ }\bibfield
  {title} {\bibinfo {title} {Measurement-induced steering of quantum systems},\
  }\href {https://doi.org/10.1103/PhysRevResearch.2.033347} {\bibfield
  {journal} {\bibinfo  {journal} {Phys. Rev. Research}\ }\textbf {\bibinfo
  {volume} {2}},\ \bibinfo {pages} {033347} (\bibinfo {year}
  {2020})}\BibitemShut {NoStop}%
\bibitem [{\citenamefont {Zabalo}\ \emph {et~al.}(2020)\citenamefont {Zabalo},
  \citenamefont {Gullans}, \citenamefont {Wilson}, \citenamefont
  {Gopalakrishnan}, \citenamefont {Huse},\ and\ \citenamefont
  {Pixley}}]{Zabalo2020}%
  \BibitemOpen
  \bibfield  {author} {\bibinfo {author} {\bibfnamefont {A.}~\bibnamefont
  {Zabalo}}, \bibinfo {author} {\bibfnamefont {M.~J.}\ \bibnamefont {Gullans}},
  \bibinfo {author} {\bibfnamefont {J.~H.}\ \bibnamefont {Wilson}}, \bibinfo
  {author} {\bibfnamefont {S.}~\bibnamefont {Gopalakrishnan}}, \bibinfo
  {author} {\bibfnamefont {D.~A.}\ \bibnamefont {Huse}},\ and\ \bibinfo
  {author} {\bibfnamefont {J.~H.}\ \bibnamefont {Pixley}},\ }\bibfield  {title}
  {\bibinfo {title} {Critical properties of the measurement-induced transition
  in random quantum circuits},\ }\href
  {https://doi.org/10.1103/PhysRevB.101.060301} {\bibfield  {journal} {\bibinfo
   {journal} {Phys. Rev. B}\ }\textbf {\bibinfo {volume} {101}},\ \bibinfo
  {pages} {060301(R)} (\bibinfo {year} {2020})}\BibitemShut {NoStop}%
\bibitem [{\citenamefont {Zhang}\ \emph {et~al.}(2020)\citenamefont {Zhang},
  \citenamefont {Reyes}, \citenamefont {Kourtis}, \citenamefont {Chamon},
  \citenamefont {Mucciolo},\ and\ \citenamefont {Ruckenstein}}]{Zhang2020}%
  \BibitemOpen
  \bibfield  {author} {\bibinfo {author} {\bibfnamefont {L.}~\bibnamefont
  {Zhang}}, \bibinfo {author} {\bibfnamefont {J.~A.}\ \bibnamefont {Reyes}},
  \bibinfo {author} {\bibfnamefont {S.}~\bibnamefont {Kourtis}}, \bibinfo
  {author} {\bibfnamefont {C.}~\bibnamefont {Chamon}}, \bibinfo {author}
  {\bibfnamefont {E.~R.}\ \bibnamefont {Mucciolo}},\ and\ \bibinfo {author}
  {\bibfnamefont {A.~E.}\ \bibnamefont {Ruckenstein}},\ }\bibfield  {title}
  {\bibinfo {title} {Nonuniversal entanglement level statistics in
  projection-driven quantum circuits},\ }\href
  {https://doi.org/10.1103/PhysRevB.101.235104} {\bibfield  {journal} {\bibinfo
   {journal} {Phys. Rev. B}\ }\textbf {\bibinfo {volume} {101}},\ \bibinfo
  {pages} {235104} (\bibinfo {year} {2020})}\BibitemShut {NoStop}%
\bibitem [{\citenamefont {Fan}\ \emph {et~al.}(2020)\citenamefont {Fan},
  \citenamefont {Vijay}, \citenamefont {Vishwanath},\ and\ \citenamefont
  {You}}]{Fan2020}%
  \BibitemOpen
  \bibfield  {author} {\bibinfo {author} {\bibfnamefont {R.}~\bibnamefont
  {Fan}}, \bibinfo {author} {\bibfnamefont {S.}~\bibnamefont {Vijay}}, \bibinfo
  {author} {\bibfnamefont {A.}~\bibnamefont {Vishwanath}},\ and\ \bibinfo
  {author} {\bibfnamefont {Y.-Z.}\ \bibnamefont {You}},\ }\bibfield  {title}
  {\bibinfo {title} {Self-organized error correction in random unitary circuits
  with measurement},\ }\Eprint {https://arxiv.org/abs/2002.12385}
  {arXiv:2002.12385}  (\bibinfo {year} {2020})\BibitemShut {NoStop}%
\bibitem [{\citenamefont {Bao}\ \emph {et~al.}(2020)\citenamefont {Bao},
  \citenamefont {Choi},\ and\ \citenamefont {Altman}}]{Bao2020}%
  \BibitemOpen
  \bibfield  {author} {\bibinfo {author} {\bibfnamefont {Y.}~\bibnamefont
  {Bao}}, \bibinfo {author} {\bibfnamefont {S.}~\bibnamefont {Choi}},\ and\
  \bibinfo {author} {\bibfnamefont {E.}~\bibnamefont {Altman}},\ }\bibfield
  {title} {\bibinfo {title} {Theory of the phase transition in random unitary
  circuits with measurements},\ }\href
  {https://doi.org/10.1103/PhysRevB.101.104301} {\bibfield  {journal} {\bibinfo
   {journal} {Phys. Rev. B}\ }\textbf {\bibinfo {volume} {101}},\ \bibinfo
  {pages} {104301} (\bibinfo {year} {2020})}\BibitemShut {NoStop}%
\bibitem [{\citenamefont {Bera}\ and\ \citenamefont {Roy}(2020)}]{Bera2020}%
  \BibitemOpen
  \bibfield  {author} {\bibinfo {author} {\bibfnamefont {A.}~\bibnamefont
  {Bera}}\ and\ \bibinfo {author} {\bibfnamefont {S.~S.}\ \bibnamefont {Roy}},\
  }\bibfield  {title} {\bibinfo {title} {Growth of genuine multipartite
  entanglement in random unitary circuits},\ }\Eprint
  {https://arxiv.org/abs/2003.12546} {arXiv:2003.12546}  (\bibinfo {year}
  {2020})\BibitemShut {NoStop}%
\bibitem [{\citenamefont {Lavasani}\ \emph {et~al.}(2020)\citenamefont
  {Lavasani}, \citenamefont {Alavirad},\ and\ \citenamefont
  {Barkeshli}}]{Lavasani2020}%
  \BibitemOpen
  \bibfield  {author} {\bibinfo {author} {\bibfnamefont {A.}~\bibnamefont
  {Lavasani}}, \bibinfo {author} {\bibfnamefont {Y.}~\bibnamefont {Alavirad}},\
  and\ \bibinfo {author} {\bibfnamefont {M.}~\bibnamefont {Barkeshli}},\
  }\bibfield  {title} {\bibinfo {title} {Measurement-induced topological
  entanglement transitions in symmetric random quantum circuits},\ }\Eprint
  {https://arxiv.org/abs/2004.07243} {arXiv:2004.07243}  (\bibinfo {year}
  {2020})\BibitemShut {NoStop}%
\bibitem [{\citenamefont {Sang}\ and\ \citenamefont {Hsieh}(2020)}]{Sang2020}%
  \BibitemOpen
  \bibfield  {author} {\bibinfo {author} {\bibfnamefont {S.}~\bibnamefont
  {Sang}}\ and\ \bibinfo {author} {\bibfnamefont {T.~H.}\ \bibnamefont
  {Hsieh}},\ }\bibfield  {title} {\bibinfo {title} {Measurement protected
  quantum phases},\ }\Eprint {https://arxiv.org/abs/2004.09509}
  {arXiv:2004.09509}  (\bibinfo {year} {2020})\BibitemShut {NoStop}%
\bibitem [{\citenamefont {Ippoliti}\ \emph {et~al.}(2020)\citenamefont
  {Ippoliti}, \citenamefont {Gullans}, \citenamefont {Gopalakrishnan},
  \citenamefont {Huse},\ and\ \citenamefont {Khemani}}]{Ippoliti2020}%
  \BibitemOpen
  \bibfield  {author} {\bibinfo {author} {\bibfnamefont {M.}~\bibnamefont
  {Ippoliti}}, \bibinfo {author} {\bibfnamefont {M.~J.}\ \bibnamefont
  {Gullans}}, \bibinfo {author} {\bibfnamefont {S.}~\bibnamefont
  {Gopalakrishnan}}, \bibinfo {author} {\bibfnamefont {D.~A.}\ \bibnamefont
  {Huse}},\ and\ \bibinfo {author} {\bibfnamefont {V.}~\bibnamefont
  {Khemani}},\ }\bibfield  {title} {\bibinfo {title} {Entanglement phase
  transitions in measurement-only dynamics},\ }\Eprint
  {https://arxiv.org/abs/2004.09560} {arXiv:2004.09560}  (\bibinfo {year}
  {2020})\BibitemShut {NoStop}%
\bibitem [{\citenamefont {Tang}\ and\ \citenamefont {Zhu}(2020)}]{Tang2020}%
  \BibitemOpen
  \bibfield  {author} {\bibinfo {author} {\bibfnamefont {Q.}~\bibnamefont
  {Tang}}\ and\ \bibinfo {author} {\bibfnamefont {W.}~\bibnamefont {Zhu}},\
  }\bibfield  {title} {\bibinfo {title} {Measurement-induced phase transition:
  A case study in the nonintegrable model by density-matrix renormalization
  group calculations},\ }\href
  {https://doi.org/10.1103/PhysRevResearch.2.013022} {\bibfield  {journal}
  {\bibinfo  {journal} {Phys. Rev. Research}\ }\textbf {\bibinfo {volume}
  {2}},\ \bibinfo {pages} {013022} (\bibinfo {year} {2020})}\BibitemShut
  {NoStop}%
\bibitem [{\citenamefont {Rossini}\ and\ \citenamefont
  {Vicari}(2020)}]{Rossini2020}%
  \BibitemOpen
  \bibfield  {author} {\bibinfo {author} {\bibfnamefont {D.}~\bibnamefont
  {Rossini}}\ and\ \bibinfo {author} {\bibfnamefont {E.}~\bibnamefont
  {Vicari}},\ }\bibfield  {title} {\bibinfo {title} {Measurement-induced
  dynamics of many-body systems at quantum criticality},\ }\href
  {https://doi.org/10.1103/PhysRevB.102.035119} {\bibfield  {journal} {\bibinfo
   {journal} {Phys. Rev. B}\ }\textbf {\bibinfo {volume} {102}},\ \bibinfo
  {pages} {035119} (\bibinfo {year} {2020})}\BibitemShut {NoStop}%
\bibitem [{\citenamefont {Goto}\ and\ \citenamefont
  {Danshita}(2020)}]{Goto2020}%
  \BibitemOpen
  \bibfield  {author} {\bibinfo {author} {\bibfnamefont {S.}~\bibnamefont
  {Goto}}\ and\ \bibinfo {author} {\bibfnamefont {I.}~\bibnamefont
  {Danshita}},\ }\bibfield  {title} {\bibinfo {title} {Measurement-induced
  transitions of the entanglement scaling law in ultracold gases with
  controllable dissipation},\ }\href
  {https://doi.org/10.1103/PhysRevA.102.033316} {\bibfield  {journal} {\bibinfo
   {journal} {Phys. Rev. A}\ }\textbf {\bibinfo {volume} {102}},\ \bibinfo
  {pages} {033316} (\bibinfo {year} {2020})}\BibitemShut {NoStop}%
\bibitem [{\citenamefont {Fuji}\ and\ \citenamefont {Ashida}(2020)}]{Fuji2020}%
  \BibitemOpen
  \bibfield  {author} {\bibinfo {author} {\bibfnamefont {Y.}~\bibnamefont
  {Fuji}}\ and\ \bibinfo {author} {\bibfnamefont {Y.}~\bibnamefont {Ashida}},\
  }\bibfield  {title} {\bibinfo {title} {Measurement-induced quantum
  criticality under continuous monitoring},\ }\href
  {https://doi.org/10.1103/PhysRevB.102.054302} {\bibfield  {journal} {\bibinfo
   {journal} {Phys. Rev. B}\ }\textbf {\bibinfo {volume} {102}},\ \bibinfo
  {pages} {054302} (\bibinfo {year} {2020})}\BibitemShut {NoStop}%
\bibitem [{\citenamefont {Snizhko}\ \emph {et~al.}(2020)\citenamefont
  {Snizhko}, \citenamefont {Kumar},\ and\ \citenamefont
  {Romito}}]{Snizhko2020}%
  \BibitemOpen
  \bibfield  {author} {\bibinfo {author} {\bibfnamefont {K.}~\bibnamefont
  {Snizhko}}, \bibinfo {author} {\bibfnamefont {P.}~\bibnamefont {Kumar}},\
  and\ \bibinfo {author} {\bibfnamefont {A.}~\bibnamefont {Romito}},\
  }\bibfield  {title} {\bibinfo {title} {Quantum zeno effect appears in
  stages},\ }\href {https://doi.org/10.1103/PhysRevResearch.2.033512}
  {\bibfield  {journal} {\bibinfo  {journal} {Phys. Rev. Research}\ }\textbf
  {\bibinfo {volume} {2}},\ \bibinfo {pages} {033512} (\bibinfo {year}
  {2020})}\BibitemShut {NoStop}%
\bibitem [{\citenamefont {Gebhart}\ \emph {et~al.}(2020)\citenamefont
  {Gebhart}, \citenamefont {Snizhko}, \citenamefont {Wellens}, \citenamefont
  {Buchleitner}, \citenamefont {Romito},\ and\ \citenamefont
  {Gefen}}]{Gebhart2020}%
  \BibitemOpen
  \bibfield  {author} {\bibinfo {author} {\bibfnamefont {V.}~\bibnamefont
  {Gebhart}}, \bibinfo {author} {\bibfnamefont {K.}~\bibnamefont {Snizhko}},
  \bibinfo {author} {\bibfnamefont {T.}~\bibnamefont {Wellens}}, \bibinfo
  {author} {\bibfnamefont {A.}~\bibnamefont {Buchleitner}}, \bibinfo {author}
  {\bibfnamefont {A.}~\bibnamefont {Romito}},\ and\ \bibinfo {author}
  {\bibfnamefont {Y.}~\bibnamefont {Gefen}},\ }\bibfield  {title} {\bibinfo
  {title} {Topological transition in measurement-induced geometric phases},\
  }\href {https://doi.org/10.1073/pnas.1911620117} {\bibfield  {journal}
  {\bibinfo  {journal} {PNAS}\ }\textbf {\bibinfo {volume} {117}},\ \bibinfo
  {pages} {5706} (\bibinfo {year} {2020})}\BibitemShut {NoStop}%
\bibitem [{\citenamefont {Gullans}\ and\ \citenamefont
  {Huse}(2020{\natexlab{b}})}]{Gullans2019}%
  \BibitemOpen
  \bibfield  {author} {\bibinfo {author} {\bibfnamefont {M.~J.}\ \bibnamefont
  {Gullans}}\ and\ \bibinfo {author} {\bibfnamefont {D.~A.}\ \bibnamefont
  {Huse}},\ }\bibfield  {title} {\bibinfo {title} {Scalable probes of
  measurement-induced criticality},\ }\href
  {https://doi.org/10.1103/PhysRevLett.125.070606} {\bibfield  {journal}
  {\bibinfo  {journal} {Phys. Rev. Lett.}\ }\textbf {\bibinfo {volume} {125}},\
  \bibinfo {pages} {070606} (\bibinfo {year} {2020}{\natexlab{b}})}\BibitemShut
  {NoStop}%
\bibitem [{\citenamefont {Lopez-Piqueres}\ \emph {et~al.}(2020)\citenamefont
  {Lopez-Piqueres}, \citenamefont {Ware},\ and\ \citenamefont
  {Vasseur}}]{LopezPiqueres2020}%
  \BibitemOpen
  \bibfield  {author} {\bibinfo {author} {\bibfnamefont {J.}~\bibnamefont
  {Lopez-Piqueres}}, \bibinfo {author} {\bibfnamefont {B.}~\bibnamefont
  {Ware}},\ and\ \bibinfo {author} {\bibfnamefont {R.}~\bibnamefont
  {Vasseur}},\ }\bibfield  {title} {\bibinfo {title} {Mean-field entanglement
  transitions in random tree tensor networks},\ }\href
  {https://doi.org/10.1103/PhysRevB.102.064202} {\bibfield  {journal} {\bibinfo
   {journal} {Phys. Rev. B}\ }\textbf {\bibinfo {volume} {102}},\ \bibinfo
  {pages} {064202} (\bibinfo {year} {2020})}\BibitemShut {NoStop}%
\bibitem [{\citenamefont {Shtanko}\ \emph {et~al.}(2020)\citenamefont
  {Shtanko}, \citenamefont {Kharkov}, \citenamefont {Garc\'ia-Pintos},\ and\
  \citenamefont {Gorshkov}}]{Shtanko2020}%
  \BibitemOpen
  \bibfield  {author} {\bibinfo {author} {\bibfnamefont {O.}~\bibnamefont
  {Shtanko}}, \bibinfo {author} {\bibfnamefont {Y.~A.}\ \bibnamefont
  {Kharkov}}, \bibinfo {author} {\bibfnamefont {L.~P.}\ \bibnamefont
  {Garc\'ia-Pintos}},\ and\ \bibinfo {author} {\bibfnamefont {A.~V.}\
  \bibnamefont {Gorshkov}},\ }\bibfield  {title} {\bibinfo {title} {Classical
  models of entanglement in monitored random circuits},\ }\Eprint
  {https://arxiv.org/abs/2004.06736} {arXiv:2004.06736}  (\bibinfo {year}
  {2020})\BibitemShut {NoStop}%
\bibitem [{\citenamefont {Noh}\ \emph {et~al.}(2020)\citenamefont {Noh},
  \citenamefont {Jiang},\ and\ \citenamefont {Fefferman}}]{Noh2020}%
  \BibitemOpen
  \bibfield  {author} {\bibinfo {author} {\bibfnamefont {K.}~\bibnamefont
  {Noh}}, \bibinfo {author} {\bibfnamefont {L.}~\bibnamefont {Jiang}},\ and\
  \bibinfo {author} {\bibfnamefont {B.}~\bibnamefont {Fefferman}},\ }\bibfield
  {title} {\bibinfo {title} {Efficient classical simulation of noisy random
  quantum circuits in one dimension},\ }\href
  {https://doi.org/10.22331/q-2020-09-11-318} {\bibfield  {journal} {\bibinfo
  {journal} {{Quantum}}\ }\textbf {\bibinfo {volume} {4}},\ \bibinfo {pages}
  {318} (\bibinfo {year} {2020})}\BibitemShut {NoStop}%
\bibitem [{\citenamefont {Fang}\ \emph {et~al.}(2020)\citenamefont {Fang},
  \citenamefont {Chen}, \citenamefont {Chen},\ and\ \citenamefont
  {Ye}}]{Fang2020}%
  \BibitemOpen
  \bibfield  {author} {\bibinfo {author} {\bibfnamefont {B.-L.}\ \bibnamefont
  {Fang}}, \bibinfo {author} {\bibfnamefont {J.}~\bibnamefont {Chen}}, \bibinfo
  {author} {\bibfnamefont {F.}~\bibnamefont {Chen}},\ and\ \bibinfo {author}
  {\bibfnamefont {L.}~\bibnamefont {Ye}},\ }\bibfield  {title} {\bibinfo
  {title} {Weak measurement induced asymmetric quantum cloning},\ }\href
  {https://doi.org/10.1088/1612-202X/ab9901} {\bibfield  {journal} {\bibinfo
  {journal} {Laser Phys. Lett.}\ }\textbf {\bibinfo {volume} {17}},\ \bibinfo
  {pages} {085203} (\bibinfo {year} {2020})}\BibitemShut {NoStop}%
\bibitem [{\citenamefont {Alberton}\ \emph {et~al.}(2020)\citenamefont
  {Alberton}, \citenamefont {Buchhold},\ and\ \citenamefont
  {Diehl}}]{Alberton2020}%
  \BibitemOpen
  \bibfield  {author} {\bibinfo {author} {\bibfnamefont {O.}~\bibnamefont
  {Alberton}}, \bibinfo {author} {\bibfnamefont {M.}~\bibnamefont {Buchhold}},\
  and\ \bibinfo {author} {\bibfnamefont {S.}~\bibnamefont {Diehl}},\ }\bibfield
   {title} {\bibinfo {title} {Trajectory dependent entanglement transition in a
  free fermion chain -- from extended criticality to area law},\ }\Eprint
  {https://arxiv.org/abs/2005.09722} {arXiv:2005.09722}  (\bibinfo {year}
  {2020})\BibitemShut {NoStop}%
\bibitem [{\citenamefont {Lunt}\ and\ \citenamefont {Pal}(2020)}]{Lunt2020}%
  \BibitemOpen
  \bibfield  {author} {\bibinfo {author} {\bibfnamefont {O.}~\bibnamefont
  {Lunt}}\ and\ \bibinfo {author} {\bibfnamefont {A.}~\bibnamefont {Pal}},\
  }\bibfield  {title} {\bibinfo {title} {Measurement-induced entanglement
  transitions in many-body localized systems},\ }\href
  {https://doi.org/10.1103/PhysRevResearch.2.043072} {\bibfield  {journal}
  {\bibinfo  {journal} {Phys. Rev. Research}\ }\textbf {\bibinfo {volume}
  {2}},\ \bibinfo {pages} {043072} (\bibinfo {year} {2020})}\BibitemShut
  {NoStop}%
\bibitem [{\citenamefont {Turkeshi}\ \emph {et~al.}(2020)\citenamefont
  {Turkeshi}, \citenamefont {Fazio},\ and\ \citenamefont
  {Dalmonte}}]{Turkeshi2020}%
  \BibitemOpen
  \bibfield  {author} {\bibinfo {author} {\bibfnamefont {X.}~\bibnamefont
  {Turkeshi}}, \bibinfo {author} {\bibfnamefont {R.}~\bibnamefont {Fazio}},\
  and\ \bibinfo {author} {\bibfnamefont {M.}~\bibnamefont {Dalmonte}},\
  }\bibfield  {title} {\bibinfo {title} {Measurement-induced criticality in
  $(2+1)$-dimensional hybrid quantum circuits},\ }\href
  {https://doi.org/10.1103/PhysRevB.102.014315} {\bibfield  {journal} {\bibinfo
   {journal} {Phys. Rev. B}\ }\textbf {\bibinfo {volume} {102}},\ \bibinfo
  {pages} {014315} (\bibinfo {year} {2020})}\BibitemShut {NoStop}%
\bibitem [{\citenamefont {Li}\ and\ \citenamefont
  {Fisher}(2020)}]{Li2020correcting}%
  \BibitemOpen
  \bibfield  {author} {\bibinfo {author} {\bibfnamefont {Y.}~\bibnamefont
  {Li}}\ and\ \bibinfo {author} {\bibfnamefont {M.~P.~A.}\ \bibnamefont
  {Fisher}},\ }\bibfield  {title} {\bibinfo {title} {Statistical mechanics of
  quantum error-correcting codes},\ }\Eprint {https://arxiv.org/abs/2007.03822}
  {arXiv:2007.03822}  (\bibinfo {year} {2020})\BibitemShut {NoStop}%
\bibitem [{\citenamefont {\ifmmode \check{Z}\else
  \v{Z}\fi{}nidari\ifmmode~\check{c}\else \v{c}\fi{}}\ \emph
  {et~al.}(2008)\citenamefont {\ifmmode \check{Z}\else
  \v{Z}\fi{}nidari\ifmmode~\check{c}\else \v{c}\fi{}}, \citenamefont {Prosen},\
  and\ \citenamefont {Prelov\ifmmode~\check{s}\else
  \v{s}\fi{}ek}}]{Znidaric2008}%
  \BibitemOpen
  \bibfield  {author} {\bibinfo {author} {\bibfnamefont {M.}~\bibnamefont
  {\ifmmode \check{Z}\else \v{Z}\fi{}nidari\ifmmode~\check{c}\else
  \v{c}\fi{}}}, \bibinfo {author} {\bibfnamefont {T.}~\bibnamefont {Prosen}},\
  and\ \bibinfo {author} {\bibfnamefont {P.}~\bibnamefont
  {Prelov\ifmmode~\check{s}\else \v{s}\fi{}ek}},\ }\bibfield  {title} {\bibinfo
  {title} {Many-body localization in the {Heisenberg} {$XXZ$} magnet in a
  random field},\ }\href {https://doi.org/10.1103/PhysRevB.77.064426}
  {\bibfield  {journal} {\bibinfo  {journal} {Phys. Rev. B}\ }\textbf {\bibinfo
  {volume} {77}},\ \bibinfo {pages} {064426} (\bibinfo {year}
  {2008})}\BibitemShut {NoStop}%
\bibitem [{\citenamefont {Eisert}\ \emph {et~al.}(2010)\citenamefont {Eisert},
  \citenamefont {Cramer},\ and\ \citenamefont {Plenio}}]{Eisert2010}%
  \BibitemOpen
  \bibfield  {author} {\bibinfo {author} {\bibfnamefont {J.}~\bibnamefont
  {Eisert}}, \bibinfo {author} {\bibfnamefont {M.}~\bibnamefont {Cramer}},\
  and\ \bibinfo {author} {\bibfnamefont {M.~B.}\ \bibnamefont {Plenio}},\
  }\bibfield  {title} {\bibinfo {title} {Colloquium: Area laws for the
  entanglement entropy},\ }\href {https://doi.org/10.1103/RevModPhys.82.277}
  {\bibfield  {journal} {\bibinfo  {journal} {Rev. Mod. Phys.}\ }\textbf
  {\bibinfo {volume} {82}},\ \bibinfo {pages} {277} (\bibinfo {year}
  {2010})}\BibitemShut {NoStop}%
\bibitem [{\citenamefont {Bardarson}\ \emph {et~al.}(2012)\citenamefont
  {Bardarson}, \citenamefont {Pollmann},\ and\ \citenamefont
  {Moore}}]{Bardarson2012}%
  \BibitemOpen
  \bibfield  {author} {\bibinfo {author} {\bibfnamefont {J.~H.}\ \bibnamefont
  {Bardarson}}, \bibinfo {author} {\bibfnamefont {F.}~\bibnamefont
  {Pollmann}},\ and\ \bibinfo {author} {\bibfnamefont {J.~E.}\ \bibnamefont
  {Moore}},\ }\bibfield  {title} {\bibinfo {title} {Unbounded growth of
  entanglement in models of many-body localization},\ }\href
  {https://doi.org/10.1103/PhysRevLett.109.017202} {\bibfield  {journal}
  {\bibinfo  {journal} {Phys. Rev. Lett.}\ }\textbf {\bibinfo {volume} {109}},\
  \bibinfo {pages} {017202} (\bibinfo {year} {2012})}\BibitemShut {NoStop}%
\bibitem [{\citenamefont {Bauer}\ and\ \citenamefont
  {Nayak}(2013)}]{Bauer2013}%
  \BibitemOpen
  \bibfield  {author} {\bibinfo {author} {\bibfnamefont {B.}~\bibnamefont
  {Bauer}}\ and\ \bibinfo {author} {\bibfnamefont {C.}~\bibnamefont {Nayak}},\
  }\bibfield  {title} {\bibinfo {title} {Area laws in a many-body localized
  state and its implications for topological order},\ }\href
  {https://doi.org/10.1088/1742-5468/2013/09/p09005} {\bibfield  {journal}
  {\bibinfo  {journal} {J. Stat. Mech. Theory Exp.}\ }\textbf {\bibinfo
  {volume} {2013}},\ \bibinfo {pages} {P09005} (\bibinfo {year}
  {2013})}\BibitemShut {NoStop}%
\bibitem [{\citenamefont {Kj\"all}\ \emph {et~al.}(2014)\citenamefont
  {Kj\"all}, \citenamefont {Bardarson},\ and\ \citenamefont
  {Pollmann}}]{Kjall2014}%
  \BibitemOpen
  \bibfield  {author} {\bibinfo {author} {\bibfnamefont {J.~A.}\ \bibnamefont
  {Kj\"all}}, \bibinfo {author} {\bibfnamefont {J.~H.}\ \bibnamefont
  {Bardarson}},\ and\ \bibinfo {author} {\bibfnamefont {F.}~\bibnamefont
  {Pollmann}},\ }\bibfield  {title} {\bibinfo {title} {Many-body localization
  in a disordered quantum {Ising} chain},\ }\href
  {https://doi.org/10.1103/PhysRevLett.113.107204} {\bibfield  {journal}
  {\bibinfo  {journal} {Phys. Rev. Lett.}\ }\textbf {\bibinfo {volume} {113}},\
  \bibinfo {pages} {107204} (\bibinfo {year} {2014})}\BibitemShut {NoStop}%
\bibitem [{\citenamefont {Luitz}\ \emph {et~al.}(2015)\citenamefont {Luitz},
  \citenamefont {Laflorencie},\ and\ \citenamefont {Alet}}]{Luitz2015}%
  \BibitemOpen
  \bibfield  {author} {\bibinfo {author} {\bibfnamefont {D.~J.}\ \bibnamefont
  {Luitz}}, \bibinfo {author} {\bibfnamefont {N.}~\bibnamefont {Laflorencie}},\
  and\ \bibinfo {author} {\bibfnamefont {F.}~\bibnamefont {Alet}},\ }\bibfield
  {title} {\bibinfo {title} {Many-body localization edge in the random-field
  {Heisenberg} chain},\ }\href {https://doi.org/10.1103/PhysRevB.91.081103}
  {\bibfield  {journal} {\bibinfo  {journal} {Phys. Rev. B}\ }\textbf {\bibinfo
  {volume} {91}},\ \bibinfo {pages} {081103(R)} (\bibinfo {year}
  {2015})}\BibitemShut {NoStop}%
\bibitem [{\citenamefont {Petta}\ \emph {et~al.}(2005)\citenamefont {Petta},
  \citenamefont {Johnson}, \citenamefont {Taylor}, \citenamefont {Laird},
  \citenamefont {Yacoby}, \citenamefont {Lukin}, \citenamefont {Marcus},
  \citenamefont {Hanson},\ and\ \citenamefont {Gossard}}]{petta2005coherent}%
  \BibitemOpen
  \bibfield  {author} {\bibinfo {author} {\bibfnamefont {J.~R.}\ \bibnamefont
  {Petta}}, \bibinfo {author} {\bibfnamefont {A.~C.}\ \bibnamefont {Johnson}},
  \bibinfo {author} {\bibfnamefont {J.~M.}\ \bibnamefont {Taylor}}, \bibinfo
  {author} {\bibfnamefont {E.~A.}\ \bibnamefont {Laird}}, \bibinfo {author}
  {\bibfnamefont {A.}~\bibnamefont {Yacoby}}, \bibinfo {author} {\bibfnamefont
  {M.~D.}\ \bibnamefont {Lukin}}, \bibinfo {author} {\bibfnamefont {C.~M.}\
  \bibnamefont {Marcus}}, \bibinfo {author} {\bibfnamefont {M.~P.}\
  \bibnamefont {Hanson}},\ and\ \bibinfo {author} {\bibfnamefont {A.~C.}\
  \bibnamefont {Gossard}},\ }\bibfield  {title} {\bibinfo {title} {Coherent
  manipulation of coupled electron spins in semiconductor quantum dots},\
  }\href {https://doi.org/10.1126/science.1116955} {\bibfield  {journal}
  {\bibinfo  {journal} {Science}\ }\textbf {\bibinfo {volume} {309}},\ \bibinfo
  {pages} {2180} (\bibinfo {year} {2005})}\BibitemShut {NoStop}%
\bibitem [{\citenamefont {Kim}\ \emph {et~al.}(2014)\citenamefont {Kim},
  \citenamefont {Shi}, \citenamefont {Simmons}, \citenamefont {Ward},
  \citenamefont {Prance}, \citenamefont {Koh}, \citenamefont {Gamble},
  \citenamefont {Savage}, \citenamefont {Lagally}, \citenamefont {Friesen}
  \emph {et~al.}}]{kim2014quantum}%
  \BibitemOpen
  \bibfield  {author} {\bibinfo {author} {\bibfnamefont {D.}~\bibnamefont
  {Kim}}, \bibinfo {author} {\bibfnamefont {Z.}~\bibnamefont {Shi}}, \bibinfo
  {author} {\bibfnamefont {C.}~\bibnamefont {Simmons}}, \bibinfo {author}
  {\bibfnamefont {D.}~\bibnamefont {Ward}}, \bibinfo {author} {\bibfnamefont
  {J.}~\bibnamefont {Prance}}, \bibinfo {author} {\bibfnamefont {T.~S.}\
  \bibnamefont {Koh}}, \bibinfo {author} {\bibfnamefont {J.~K.}\ \bibnamefont
  {Gamble}}, \bibinfo {author} {\bibfnamefont {D.}~\bibnamefont {Savage}},
  \bibinfo {author} {\bibfnamefont {M.}~\bibnamefont {Lagally}}, \bibinfo
  {author} {\bibfnamefont {M.}~\bibnamefont {Friesen}}, \emph {et~al.},\
  }\bibfield  {title} {\bibinfo {title} {Quantum control and process tomography
  of a semiconductor quantum dot hybrid qubit},\ }\href
  {https://doi.org/10.1038/nature13407} {\bibfield  {journal} {\bibinfo
  {journal} {Nature}\ }\textbf {\bibinfo {volume} {511}},\ \bibinfo {pages}
  {70} (\bibinfo {year} {2014})}\BibitemShut {NoStop}%
\bibitem [{\citenamefont {West}\ \emph {et~al.}(2019)\citenamefont {West},
  \citenamefont {Hensen}, \citenamefont {Jouan}, \citenamefont {Tanttu},
  \citenamefont {Yang}, \citenamefont {Rossi}, \citenamefont {Gonzalez-Zalba},
  \citenamefont {Hudson}, \citenamefont {Morello}, \citenamefont {Reilly} \emph
  {et~al.}}]{west2019gate}%
  \BibitemOpen
  \bibfield  {author} {\bibinfo {author} {\bibfnamefont {A.}~\bibnamefont
  {West}}, \bibinfo {author} {\bibfnamefont {B.}~\bibnamefont {Hensen}},
  \bibinfo {author} {\bibfnamefont {A.}~\bibnamefont {Jouan}}, \bibinfo
  {author} {\bibfnamefont {T.}~\bibnamefont {Tanttu}}, \bibinfo {author}
  {\bibfnamefont {C.-H.}\ \bibnamefont {Yang}}, \bibinfo {author}
  {\bibfnamefont {A.}~\bibnamefont {Rossi}}, \bibinfo {author} {\bibfnamefont
  {M.~F.}\ \bibnamefont {Gonzalez-Zalba}}, \bibinfo {author} {\bibfnamefont
  {F.}~\bibnamefont {Hudson}}, \bibinfo {author} {\bibfnamefont
  {A.}~\bibnamefont {Morello}}, \bibinfo {author} {\bibfnamefont {D.~J.}\
  \bibnamefont {Reilly}}, \emph {et~al.},\ }\bibfield  {title} {\bibinfo
  {title} {Gate-based single-shot readout of spins in silicon},\ }\href
  {https://doi.org/10.1038/s41565-019-0400-7} {\bibfield  {journal} {\bibinfo
  {journal} {Nat. Nanotechnol.}\ }\textbf {\bibinfo {volume} {14}},\ \bibinfo
  {pages} {437} (\bibinfo {year} {2019})}\BibitemShut {NoStop}%
\bibitem [{\citenamefont {Murch}\ \emph {et~al.}(2013)\citenamefont {Murch},
  \citenamefont {Weber}, \citenamefont {Macklin},\ and\ \citenamefont
  {Siddiqi}}]{Murch2013observing}%
  \BibitemOpen
  \bibfield  {author} {\bibinfo {author} {\bibfnamefont {K.}~\bibnamefont
  {Murch}}, \bibinfo {author} {\bibfnamefont {S.}~\bibnamefont {Weber}},
  \bibinfo {author} {\bibfnamefont {C.}~\bibnamefont {Macklin}},\ and\ \bibinfo
  {author} {\bibfnamefont {I.}~\bibnamefont {Siddiqi}},\ }\bibfield  {title}
  {\bibinfo {title} {Observing single quantum trajectories of a superconducting
  quantum bit},\ }\href {https://doi.org/10.1038/nature12539} {\bibfield
  {journal} {\bibinfo  {journal} {Nature}\ }\textbf {\bibinfo {volume} {502}},\
  \bibinfo {pages} {211} (\bibinfo {year} {2013})}\BibitemShut {NoStop}%
\bibitem [{\citenamefont {Neumann}\ \emph {et~al.}(2010)\citenamefont
  {Neumann}, \citenamefont {Beck}, \citenamefont {Steiner}, \citenamefont
  {Rempp}, \citenamefont {Fedder}, \citenamefont {Hemmer}, \citenamefont
  {Wrachtrup},\ and\ \citenamefont {Jelezko}}]{neumann2010single}%
  \BibitemOpen
  \bibfield  {author} {\bibinfo {author} {\bibfnamefont {P.}~\bibnamefont
  {Neumann}}, \bibinfo {author} {\bibfnamefont {J.}~\bibnamefont {Beck}},
  \bibinfo {author} {\bibfnamefont {M.}~\bibnamefont {Steiner}}, \bibinfo
  {author} {\bibfnamefont {F.}~\bibnamefont {Rempp}}, \bibinfo {author}
  {\bibfnamefont {H.}~\bibnamefont {Fedder}}, \bibinfo {author} {\bibfnamefont
  {P.~R.}\ \bibnamefont {Hemmer}}, \bibinfo {author} {\bibfnamefont
  {J.}~\bibnamefont {Wrachtrup}},\ and\ \bibinfo {author} {\bibfnamefont
  {F.}~\bibnamefont {Jelezko}},\ }\bibfield  {title} {\bibinfo {title}
  {Single-shot readout of a single nuclear spin},\ }\href
  {https://doi.org/10.1126/science.1189075} {\bibfield  {journal} {\bibinfo
  {journal} {Science}\ }\textbf {\bibinfo {volume} {329}},\ \bibinfo {pages}
  {542} (\bibinfo {year} {2010})}\BibitemShut {NoStop}%
\bibitem [{\citenamefont {Aasen}\ \emph {et~al.}(2016)\citenamefont {Aasen},
  \citenamefont {Hell}, \citenamefont {Mishmash}, \citenamefont {Higginbotham},
  \citenamefont {Danon}, \citenamefont {Leijnse}, \citenamefont {Jespersen},
  \citenamefont {Folk}, \citenamefont {Marcus}, \citenamefont {Flensberg},\
  and\ \citenamefont {Alicea}}]{Aasen2016milestones}%
  \BibitemOpen
  \bibfield  {author} {\bibinfo {author} {\bibfnamefont {D.}~\bibnamefont
  {Aasen}}, \bibinfo {author} {\bibfnamefont {M.}~\bibnamefont {Hell}},
  \bibinfo {author} {\bibfnamefont {R.~V.}\ \bibnamefont {Mishmash}}, \bibinfo
  {author} {\bibfnamefont {A.}~\bibnamefont {Higginbotham}}, \bibinfo {author}
  {\bibfnamefont {J.}~\bibnamefont {Danon}}, \bibinfo {author} {\bibfnamefont
  {M.}~\bibnamefont {Leijnse}}, \bibinfo {author} {\bibfnamefont {T.~S.}\
  \bibnamefont {Jespersen}}, \bibinfo {author} {\bibfnamefont {J.~A.}\
  \bibnamefont {Folk}}, \bibinfo {author} {\bibfnamefont {C.~M.}\ \bibnamefont
  {Marcus}}, \bibinfo {author} {\bibfnamefont {K.}~\bibnamefont {Flensberg}},\
  and\ \bibinfo {author} {\bibfnamefont {J.}~\bibnamefont {Alicea}},\
  }\bibfield  {title} {\bibinfo {title} {Milestones toward {Majorana}-based
  quantum computing},\ }\href {https://doi.org/10.1103/PhysRevX.6.031016}
  {\bibfield  {journal} {\bibinfo  {journal} {Phys. Rev. X}\ }\textbf {\bibinfo
  {volume} {6}},\ \bibinfo {pages} {031016} (\bibinfo {year}
  {2016})}\BibitemShut {NoStop}%
\bibitem [{\citenamefont {Manousakis}\ \emph {et~al.}(2020)\citenamefont
  {Manousakis}, \citenamefont {Wille}, \citenamefont {Altland}, \citenamefont
  {Egger}, \citenamefont {Flensberg},\ and\ \citenamefont
  {Hassler}}]{manousakis2020weak}%
  \BibitemOpen
  \bibfield  {author} {\bibinfo {author} {\bibfnamefont {J.}~\bibnamefont
  {Manousakis}}, \bibinfo {author} {\bibfnamefont {C.}~\bibnamefont {Wille}},
  \bibinfo {author} {\bibfnamefont {A.}~\bibnamefont {Altland}}, \bibinfo
  {author} {\bibfnamefont {R.}~\bibnamefont {Egger}}, \bibinfo {author}
  {\bibfnamefont {K.}~\bibnamefont {Flensberg}},\ and\ \bibinfo {author}
  {\bibfnamefont {F.}~\bibnamefont {Hassler}},\ }\bibfield  {title} {\bibinfo
  {title} {Weak measurement protocols for {Majorana} bound state
  identification},\ }\href {https://doi.org/10.1103/PhysRevLett.124.096801}
  {\bibfield  {journal} {\bibinfo  {journal} {Phys. Rev. Lett.}\ }\textbf
  {\bibinfo {volume} {124}},\ \bibinfo {pages} {096801} (\bibinfo {year}
  {2020})}\BibitemShut {NoStop}%
\bibitem [{\citenamefont {Steiner}\ and\ \citenamefont {von
  Oppen}(2020)}]{steiner2020readout}%
  \BibitemOpen
  \bibfield  {author} {\bibinfo {author} {\bibfnamefont {J.~F.}\ \bibnamefont
  {Steiner}}\ and\ \bibinfo {author} {\bibfnamefont {F.}~\bibnamefont {von
  Oppen}},\ }\bibfield  {title} {\bibinfo {title} {Readout of majorana
  qubits},\ }\href {https://doi.org/10.1103/PhysRevResearch.2.033255}
  {\bibfield  {journal} {\bibinfo  {journal} {Phys. Rev. Research}\ }\textbf
  {\bibinfo {volume} {2}},\ \bibinfo {pages} {033255} (\bibinfo {year}
  {2020})}\BibitemShut {NoStop}%
\bibitem [{\citenamefont {Munk}\ \emph {et~al.}(2020)\citenamefont {Munk},
  \citenamefont {Schulenborg}, \citenamefont {Egger},\ and\ \citenamefont
  {Flensberg}}]{munk2020parity}%
  \BibitemOpen
  \bibfield  {author} {\bibinfo {author} {\bibfnamefont {M.~I.~K.}\
  \bibnamefont {Munk}}, \bibinfo {author} {\bibfnamefont {J.}~\bibnamefont
  {Schulenborg}}, \bibinfo {author} {\bibfnamefont {R.}~\bibnamefont {Egger}},\
  and\ \bibinfo {author} {\bibfnamefont {K.}~\bibnamefont {Flensberg}},\
  }\bibfield  {title} {\bibinfo {title} {Parity-to-charge conversion in
  majorana qubit readout},\ }\href
  {https://doi.org/10.1103/PhysRevResearch.2.033254} {\bibfield  {journal}
  {\bibinfo  {journal} {Phys. Rev. Research}\ }\textbf {\bibinfo {volume}
  {2}},\ \bibinfo {pages} {033254} (\bibinfo {year} {2020})}\BibitemShut
  {NoStop}%
\bibitem [{\citenamefont {Maioli}\ \emph {et~al.}(2005)\citenamefont {Maioli},
  \citenamefont {Meunier}, \citenamefont {Gleyzes}, \citenamefont {Auffeves},
  \citenamefont {Nogues}, \citenamefont {Brune}, \citenamefont {Raimond},\ and\
  \citenamefont {Haroche}}]{maioli2005nondestructive}%
  \BibitemOpen
  \bibfield  {author} {\bibinfo {author} {\bibfnamefont {P.}~\bibnamefont
  {Maioli}}, \bibinfo {author} {\bibfnamefont {T.}~\bibnamefont {Meunier}},
  \bibinfo {author} {\bibfnamefont {S.}~\bibnamefont {Gleyzes}}, \bibinfo
  {author} {\bibfnamefont {A.}~\bibnamefont {Auffeves}}, \bibinfo {author}
  {\bibfnamefont {G.}~\bibnamefont {Nogues}}, \bibinfo {author} {\bibfnamefont
  {M.}~\bibnamefont {Brune}}, \bibinfo {author} {\bibfnamefont {J.~M.}\
  \bibnamefont {Raimond}},\ and\ \bibinfo {author} {\bibfnamefont
  {S.}~\bibnamefont {Haroche}},\ }\bibfield  {title} {\bibinfo {title}
  {{Nondestructive Rydberg atom counting with mesoscopic fields in a cavity}},\
  }\href {https://doi.org/10.1103/PhysRevLett.94.113601} {\bibfield  {journal}
  {\bibinfo  {journal} {Phys. Rev. Lett.}\ }\textbf {\bibinfo {volume} {94}},\
  \bibinfo {pages} {113601} (\bibinfo {year} {2005})}\BibitemShut {NoStop}%
\bibitem [{\citenamefont {Cao}\ \emph {et~al.}(2019)\citenamefont {Cao},
  \citenamefont {Tilloy},\ and\ \citenamefont {Luca}}]{Cao2019}%
  \BibitemOpen
  \bibfield  {author} {\bibinfo {author} {\bibfnamefont {X.}~\bibnamefont
  {Cao}}, \bibinfo {author} {\bibfnamefont {A.}~\bibnamefont {Tilloy}},\ and\
  \bibinfo {author} {\bibfnamefont {A.~D.}\ \bibnamefont {Luca}},\ }\bibfield
  {title} {\bibinfo {title} {{Entanglement in a fermion chain under continuous
  monitoring}},\ }\href {https://doi.org/10.21468/SciPostPhys.7.2.024}
  {\bibfield  {journal} {\bibinfo  {journal} {SciPost Phys.}\ }\textbf
  {\bibinfo {volume} {7}},\ \bibinfo {pages} {24} (\bibinfo {year}
  {2019})}\BibitemShut {NoStop}%
\bibitem [{\citenamefont {Napp}\ \emph {et~al.}(2019)\citenamefont {Napp},
  \citenamefont {Placa}, \citenamefont {Dalzell}, \citenamefont {Brandao},\
  and\ \citenamefont {Harrow}}]{Napp2019}%
  \BibitemOpen
  \bibfield  {author} {\bibinfo {author} {\bibfnamefont {J.}~\bibnamefont
  {Napp}}, \bibinfo {author} {\bibfnamefont {R.~L.~L.}\ \bibnamefont {Placa}},
  \bibinfo {author} {\bibfnamefont {A.~M.}\ \bibnamefont {Dalzell}}, \bibinfo
  {author} {\bibfnamefont {F.~G. S.~L.}\ \bibnamefont {Brandao}},\ and\
  \bibinfo {author} {\bibfnamefont {A.~W.}\ \bibnamefont {Harrow}},\ }\bibfield
   {title} {\bibinfo {title} {Efficient classical simulation of random shallow
  2d quantum circuits},\ }\Eprint {https://arxiv.org/abs/2001.00021}
  {arXiv:2001.00021}  (\bibinfo {year} {2019})\BibitemShut {NoStop}%
\bibitem [{\citenamefont {Zhou}\ and\ \citenamefont {Nahum}(2020)}]{Zhou2019}%
  \BibitemOpen
  \bibfield  {author} {\bibinfo {author} {\bibfnamefont {T.}~\bibnamefont
  {Zhou}}\ and\ \bibinfo {author} {\bibfnamefont {A.}~\bibnamefont {Nahum}},\
  }\bibfield  {title} {\bibinfo {title} {Entanglement membrane in chaotic
  many-body systems},\ }\href {https://doi.org/10.1103/PhysRevX.10.031066}
  {\bibfield  {journal} {\bibinfo  {journal} {Phys. Rev. X}\ }\textbf {\bibinfo
  {volume} {10}},\ \bibinfo {pages} {031066} (\bibinfo {year}
  {2020})}\BibitemShut {NoStop}%
\bibitem [{\citenamefont {Nahum}\ \emph {et~al.}(2017)\citenamefont {Nahum},
  \citenamefont {Ruhman}, \citenamefont {Vijay},\ and\ \citenamefont
  {Haah}}]{Nahum2017}%
  \BibitemOpen
  \bibfield  {author} {\bibinfo {author} {\bibfnamefont {A.}~\bibnamefont
  {Nahum}}, \bibinfo {author} {\bibfnamefont {J.}~\bibnamefont {Ruhman}},
  \bibinfo {author} {\bibfnamefont {S.}~\bibnamefont {Vijay}},\ and\ \bibinfo
  {author} {\bibfnamefont {J.}~\bibnamefont {Haah}},\ }\bibfield  {title}
  {\bibinfo {title} {Quantum entanglement growth under random unitary
  dynamics},\ }\href {https://doi.org/10.1103/PhysRevX.7.031016} {\bibfield
  {journal} {\bibinfo  {journal} {Phys. Rev. X}\ }\textbf {\bibinfo {volume}
  {7}},\ \bibinfo {pages} {031016} (\bibinfo {year} {2017})}\BibitemShut
  {NoStop}%
\bibitem [{\citenamefont {von Keyserlingk}\ \emph {et~al.}(2018)\citenamefont
  {von Keyserlingk}, \citenamefont {Rakovszky}, \citenamefont {Pollmann},\ and\
  \citenamefont {Sondhi}}]{Keyserlingk2018}%
  \BibitemOpen
  \bibfield  {author} {\bibinfo {author} {\bibfnamefont {C.~W.}\ \bibnamefont
  {von Keyserlingk}}, \bibinfo {author} {\bibfnamefont {T.}~\bibnamefont
  {Rakovszky}}, \bibinfo {author} {\bibfnamefont {F.}~\bibnamefont
  {Pollmann}},\ and\ \bibinfo {author} {\bibfnamefont {S.~L.}\ \bibnamefont
  {Sondhi}},\ }\bibfield  {title} {\bibinfo {title} {Operator hydrodynamics,
  {OTOCs}, and entanglement growth in systems without conservation laws},\
  }\href {https://doi.org/10.1103/PhysRevX.8.021013} {\bibfield  {journal}
  {\bibinfo  {journal} {Phys. Rev. X}\ }\textbf {\bibinfo {volume} {8}},\
  \bibinfo {pages} {021013} (\bibinfo {year} {2018})}\BibitemShut {NoStop}%
\bibitem [{\citenamefont {Nahum}\ \emph {et~al.}(2018)\citenamefont {Nahum},
  \citenamefont {Vijay},\ and\ \citenamefont {Haah}}]{Nahum2018}%
  \BibitemOpen
  \bibfield  {author} {\bibinfo {author} {\bibfnamefont {A.}~\bibnamefont
  {Nahum}}, \bibinfo {author} {\bibfnamefont {S.}~\bibnamefont {Vijay}},\ and\
  \bibinfo {author} {\bibfnamefont {J.}~\bibnamefont {Haah}},\ }\bibfield
  {title} {\bibinfo {title} {Operator spreading in random unitary circuits},\
  }\href {https://doi.org/10.1103/PhysRevX.8.021014} {\bibfield  {journal}
  {\bibinfo  {journal} {Phys. Rev. X}\ }\textbf {\bibinfo {volume} {8}},\
  \bibinfo {pages} {021014} (\bibinfo {year} {2018})}\BibitemShut {NoStop}%
\bibitem [{\citenamefont {Hua}(1963)}]{Hua1963}%
  \BibitemOpen
  \bibfield  {author} {\bibinfo {author} {\bibfnamefont {L.}~\bibnamefont
  {Hua}},\ }\href@noop {} {\emph {\bibinfo {title} {Harmonic analysis of
  functions of several complex variables in the classical domains}}}\ (\bibinfo
   {publisher} {American Mathematical Society},\ \bibinfo {address}
  {Providence, R.I.},\ \bibinfo {year} {1963})\BibitemShut {NoStop}%
\bibitem [{Note1()}]{Note1}%
  \BibitemOpen
  \bibinfo {note} {The Cauchy process arises from the Lorentzian distribution
  of $H_\protect \mathrm {eff}$; see, e.g., Ref.~\cite {Brouwer1995}. An
  analogous entanglement dynamics can be obtained from a Wiener process where
  $H_\protect \mathrm {eff}$ is taken from the Gaussian unitary ensemble with
  matrix elements $H_{\protect \mathrm {eff},{lm}}=O(1/\mu )$; for more details
  see \cite {Note2}.}\BibitemShut {Stop}%
\bibitem [{\citenamefont {Mello}\ \emph {et~al.}(1985)\citenamefont {Mello},
  \citenamefont {Pereyra},\ and\ \citenamefont {Seligman}}]{Mello1985}%
  \BibitemOpen
  \bibfield  {author} {\bibinfo {author} {\bibfnamefont {P.~A.}\ \bibnamefont
  {Mello}}, \bibinfo {author} {\bibfnamefont {P.}~\bibnamefont {Pereyra}},\
  and\ \bibinfo {author} {\bibfnamefont {T.~H.}\ \bibnamefont {Seligman}},\
  }\bibfield  {title} {\bibinfo {title} {Information theory and statistical
  nuclear reactions. i. general theory and applications to few-channel
  problems},\ }\href
  {https://doi.org/https://doi.org/10.1016/0003-4916(85)90080-6} {\bibfield
  {journal} {\bibinfo  {journal} {Ann. Phys. (N. Y.)}\ }\textbf {\bibinfo
  {volume} {161}},\ \bibinfo {pages} {254} (\bibinfo {year}
  {1985})}\BibitemShut {NoStop}%
\bibitem [{\citenamefont {Brouwer}(1995)}]{Brouwer1995}%
  \BibitemOpen
  \bibfield  {author} {\bibinfo {author} {\bibfnamefont {P.~W.}\ \bibnamefont
  {Brouwer}},\ }\bibfield  {title} {\bibinfo {title} {Generalized circular
  ensemble of scattering matrices for a chaotic cavity with nonideal leads},\
  }\href {https://doi.org/10.1103/PhysRevB.51.16878} {\bibfield  {journal}
  {\bibinfo  {journal} {Phys. Rev. B}\ }\textbf {\bibinfo {volume} {51}},\
  \bibinfo {pages} {16878} (\bibinfo {year} {1995})}\BibitemShut {NoStop}%
\bibitem [{\citenamefont {Jacobs}(2014)}]{Jacobs2014quantum}%
  \BibitemOpen
  \bibfield  {author} {\bibinfo {author} {\bibfnamefont {K.}~\bibnamefont
  {Jacobs}},\ }\href {https://doi.org/10.1017/CBO9781139179027} {\emph
  {\bibinfo {title} {Quantum measurement theory and its applications}}}\
  (\bibinfo  {publisher} {Cambridge University Press},\ \bibinfo {year}
  {2014})\BibitemShut {NoStop}%
\bibitem [{\citenamefont {Wiseman}\ and\ \citenamefont
  {Milburn}(2009)}]{Wiseman2009quantum}%
  \BibitemOpen
  \bibfield  {author} {\bibinfo {author} {\bibfnamefont {H.~M.}\ \bibnamefont
  {Wiseman}}\ and\ \bibinfo {author} {\bibfnamefont {G.~J.}\ \bibnamefont
  {Milburn}},\ }\href {https://doi.org/10.1017/CBO9780511813948} {\emph
  {\bibinfo {title} {Quantum measurement and control}}}\ (\bibinfo  {publisher}
  {Cambridge University Press},\ \bibinfo {year} {2009})\BibitemShut {NoStop}%
\bibitem [{Note2()}]{Note2}%
  \BibitemOpen
  \bibinfo {note} {See the Supplemental Material for an analysis of the
  transition in terms of the bipartite information, a discussion of the role of
  measurement frequency, the description and further discussion of the
  finite-size scaling, and a comparison between different stochastic processes
  that can be used to approach the continuum limit. This Supplemental Material
  also includes a reference to Ref. \cite {Haake1996}.}\BibitemShut {Stop}%
\bibitem [{\citenamefont {Field}\ \emph {et~al.}(1993)\citenamefont {Field},
  \citenamefont {Smith}, \citenamefont {Pepper}, \citenamefont {Ritchie},
  \citenamefont {Frost}, \citenamefont {Jones},\ and\ \citenamefont
  {Hasko}}]{Field1993measurements}%
  \BibitemOpen
  \bibfield  {author} {\bibinfo {author} {\bibfnamefont {M.}~\bibnamefont
  {Field}}, \bibinfo {author} {\bibfnamefont {C.~G.}\ \bibnamefont {Smith}},
  \bibinfo {author} {\bibfnamefont {M.}~\bibnamefont {Pepper}}, \bibinfo
  {author} {\bibfnamefont {D.~A.}\ \bibnamefont {Ritchie}}, \bibinfo {author}
  {\bibfnamefont {J.~E.~F.}\ \bibnamefont {Frost}}, \bibinfo {author}
  {\bibfnamefont {G.~A.~C.}\ \bibnamefont {Jones}},\ and\ \bibinfo {author}
  {\bibfnamefont {D.~G.}\ \bibnamefont {Hasko}},\ }\bibfield  {title} {\bibinfo
  {title} {{Measurements of Coulomb blockade with a noninvasive voltage
  probe}},\ }\href {https://doi.org/10.1103/PhysRevLett.70.1311} {\bibfield
  {journal} {\bibinfo  {journal} {Phys. Rev. Lett.}\ }\textbf {\bibinfo
  {volume} {70}},\ \bibinfo {pages} {1311} (\bibinfo {year}
  {1993})}\BibitemShut {NoStop}%
\bibitem [{\citenamefont {Korotkov}(1999)}]{Korotkov1999continuous}%
  \BibitemOpen
  \bibfield  {author} {\bibinfo {author} {\bibfnamefont {A.~N.}\ \bibnamefont
  {Korotkov}},\ }\bibfield  {title} {\bibinfo {title} {Continuous quantum
  measurement of a double dot},\ }\href
  {https://doi.org/10.1103/PhysRevB.60.5737} {\bibfield  {journal} {\bibinfo
  {journal} {Phys. Rev. B}\ }\textbf {\bibinfo {volume} {60}},\ \bibinfo
  {pages} {5737} (\bibinfo {year} {1999})}\BibitemShut {NoStop}%
\bibitem [{\citenamefont {Romito}\ \emph {et~al.}(2008)\citenamefont {Romito},
  \citenamefont {Gefen},\ and\ \citenamefont {Blanter}}]{Romito2008weak}%
  \BibitemOpen
  \bibfield  {author} {\bibinfo {author} {\bibfnamefont {A.}~\bibnamefont
  {Romito}}, \bibinfo {author} {\bibfnamefont {Y.}~\bibnamefont {Gefen}},\ and\
  \bibinfo {author} {\bibfnamefont {Y.~M.}\ \bibnamefont {Blanter}},\
  }\bibfield  {title} {\bibinfo {title} {Weak values of electron spin in a
  double quantum dot},\ }\href {https://doi.org/10.1103/PhysRevLett.100.056801}
  {\bibfield  {journal} {\bibinfo  {journal} {Phys. Rev. Lett.}\ }\textbf
  {\bibinfo {volume} {100}},\ \bibinfo {pages} {056801} (\bibinfo {year}
  {2008})}\BibitemShut {NoStop}%
\bibitem [{\citenamefont {Hosten}\ and\ \citenamefont
  {Kwiat}(2008)}]{Hosten2008observation}%
  \BibitemOpen
  \bibfield  {author} {\bibinfo {author} {\bibfnamefont {O.}~\bibnamefont
  {Hosten}}\ and\ \bibinfo {author} {\bibfnamefont {P.}~\bibnamefont {Kwiat}},\
  }\bibfield  {title} {\bibinfo {title} {{Observation of the spin Hall effect
  of light via weak measurements}},\ }\href
  {https://doi.org/10.1126/science.1152697} {\bibfield  {journal} {\bibinfo
  {journal} {Science}\ }\textbf {\bibinfo {volume} {319}},\ \bibinfo {pages}
  {787} (\bibinfo {year} {2008})}\BibitemShut {NoStop}%
\bibitem [{\citenamefont {Dixon}\ \emph {et~al.}(2009)\citenamefont {Dixon},
  \citenamefont {Starling}, \citenamefont {Jordan},\ and\ \citenamefont
  {Howell}}]{Dixon2009ultrasensitive}%
  \BibitemOpen
  \bibfield  {author} {\bibinfo {author} {\bibfnamefont {P.~B.}\ \bibnamefont
  {Dixon}}, \bibinfo {author} {\bibfnamefont {D.~J.}\ \bibnamefont {Starling}},
  \bibinfo {author} {\bibfnamefont {A.~N.}\ \bibnamefont {Jordan}},\ and\
  \bibinfo {author} {\bibfnamefont {J.~C.}\ \bibnamefont {Howell}},\ }\bibfield
   {title} {\bibinfo {title} {Ultrasensitive beam deflection measurement via
  interferometric weak value amplification},\ }\href
  {https://doi.org/10.1103/PhysRevLett.102.173601} {\bibfield  {journal}
  {\bibinfo  {journal} {Phys. Rev. Lett.}\ }\textbf {\bibinfo {volume} {102}},\
  \bibinfo {pages} {173601} (\bibinfo {year} {2009})}\BibitemShut {NoStop}%
\bibitem [{\citenamefont {Haake}\ \emph {et~al.}(1996)\citenamefont {Haake},
  \citenamefont {Kus}, \citenamefont {Sommers}, \citenamefont {Schomerus},\
  and\ \citenamefont {Zyczkowski}}]{Haake1996}%
  \BibitemOpen
  \bibfield  {author} {\bibinfo {author} {\bibfnamefont {F.}~\bibnamefont
  {Haake}}, \bibinfo {author} {\bibfnamefont {M.}~\bibnamefont {Kus}}, \bibinfo
  {author} {\bibfnamefont {H.-J.}\ \bibnamefont {Sommers}}, \bibinfo {author}
  {\bibfnamefont {H.}~\bibnamefont {Schomerus}},\ and\ \bibinfo {author}
  {\bibfnamefont {K.}~\bibnamefont {Zyczkowski}},\ }\bibfield  {title}
  {\bibinfo {title} {Secular determinants of random unitary matrices},\ }\href
  {https://doi.org/10.1088/0305-4470/29/13/029} {\bibfield  {journal} {\bibinfo
   {journal} {J. Phys. A}\ }\textbf {\bibinfo {volume} {29}},\ \bibinfo {pages}
  {3641} (\bibinfo {year} {1996})}\BibitemShut {NoStop}%
\end{thebibliography}%

\end{document}